\documentclass[12pt]{article}

\usepackage{latexsym,amsmath,amssymb,amsfonts,amsthm,bbm,mathrsfs,breakcites,dsfont,xcolor,bm}
\usepackage{stmaryrd,epsf}
\usepackage{algorithm}
\usepackage{algpseudocode}
\usepackage{bm}
\usepackage{natbib}
\usepackage{url}
\usepackage{xurl}
\usepackage{dsfont}
\usepackage{enumerate}
\usepackage{graphicx}
\usepackage{graphics}
\usepackage{psfrag}
\usepackage{caption,subcaption}
\usepackage[colorlinks,linkcolor=red,citecolor=blue,urlcolor=blue,breaklinks]{hyperref}
\usepackage{needspace}
\usepackage[export]{adjustbox}  
\usepackage{hyperref}

\newtheorem{theorem}{Theorem}

\DeclareMathOperator*{\argmin}{argmin}
\DeclareMathOperator*{\argmax}{argmax}

\newcommand{\IR}{\mbox{$\mathbb{R}$}}

\newcommand{\IE}{\mbox{$\mathbb{E}$}}

\newcommand{\KL}{\mbox{$\mathrm{KL}$}}
\newcommand{\ELBO}{\mbox{$\mathrm{ELBO}$}}

\usepackage{xcolor}
\usepackage[draft,inline,nomargin,index]{fixme}
\usepackage{placeins}

\title{\bf Scalable Bayesian Spatial Mixture Modelling\\for Remote Sensing Image Segmentation}

\author{
Bao Khanh Nguyen$^{1,}$\footnote{Corresponding author, \url{b.k.nguyen@sms.ed.ac.uk}. }, Iain Cameron$^2$, \\Cecilia Balocchi$^1$, and 
 Torben Sell$^{1,}$\footnote{The research of Torben Sell was supported by an Engineering and Physical Sciences Research Council Impact Acceleration Account award (EPSRC IAA PV185). The authors gratefully acknowledge the support of UKRI Innovate through the TAIM project. We thank the team at EOLAS Insight Ltd. for fruitful input and discussions, which helped shape the developed methodology. Author contributions are given in Section~\ref{sec:acknowledgments}.} \\[2mm]
  $^1$School of Mathematics and Maxwell Institute for Mathematical \\Sciences, University of Edinburgh, Edinburgh, UK \\
  $^2$EOLAS Insight Ltd., Edinburgh, UK
}

\date{}

\begin{document}

\maketitle

\begin{abstract}
Accurate and scalable land cover classification is essential for global conservation monitoring and policy-making.  While remote sensing images provide a cost-effective alternative to ground surveys, current methods often lack principled uncertainty quantification and require substantial labelled data, limiting their usability and reliability in new regions with distribution shifts.  We propose a Bayesian spatial mixture modelling approach for image segmentation, extending the classical Potts model by allowing for a generalised spatial dependence structure and incorporating informative priors estimated from pre-existing labelled data.  Our framework, called POTTERS (Potts Model for Enhanced Remote Sensing), enables robust uncertainty quantification, accounts for class interactions, and can detect new clusters in the target region of interest.  Crucially, our model does not require labelled data from the target region; instead, it incorporates prior information about the labels from pre-existing externally labelled images.  To ensure scalability to large remote sensing images, we develop an efficient variational inference algorithm for posterior approximation.  We demonstrate the benefits of our approach in simulation studies and apply it to land cover classification in a case study in Scotland, leveraging publicly available remote sensing data from England.  
\end{abstract}
\vspace{0.3cm}

\section{Introduction}
Increasing efforts in protecting our Earth's habitats and liveability have highlighted the need to monitor and validate habitat cover, land use, and biodiversity on a global scale \citep{abdi2020land, phiri2020sentinel,macarringue2022developments,medina2023machine}.  
In order to reduce costs and facilitate continuous validation, remote sensing imagery is increasingly used as a relatively cheap and globally scalable alternative to costly ground surveys by trained ecologists.  While remote sensing images are available in both the public and private domains at relatively low cost, there are currently severe limitations hindering their widespread adoption in policy- and decision-making.

Some major drawbacks of existing methodologies are that they require large amounts of training data, that they only provide a single estimate or inaccurate measures of uncertainty, and that they may require the user to have good domain knowledge and enough technical expertise to optimise algorithmic parameters; see~\cite{maxwell2018implementation,shirmard2022review} for two recent review articles.  Deep neural network-based methods are known to overestimate their prediction accuracy, especially in regions with little training data~\citep{kristiadi2020being}, a common scenario in remote sensing imagery.  Furthermore, algorithms trained on data from one geographical region are employed on other regions without taking into account any distribution shifts, such as, e.g., Scotland having different types of forests than England~\citep{mason2007changes}, resulting in a mean shift of the ``forest'' class between the two nations. 

We here propose a principled and rigorous methodology which allows accurate uncertainty quantification and takes into account changes in spatial and regional patterns. Our approach builds on the Ising and Potts models~\citep{cipra1987introduction,wu1982potts} to capture spatial correlation, see for example~\cite{green2002hidden, orbanz2008nonparametric, lu2020bayesian}, and see \cite{wade2023bayesian} for a review of Bayesian clustering. We generalise the aforementioned models to include spatial interactions and informed priors, which we learn from existing labelled data sets. Our approach also allows the model to find new clusters that are not available in the existing data. These contributions are then incorporated into the full posterior over all possible classification maps. As the full posterior is computationally intractable due to the large size of the remote sensing images, we propose a variational inference (VI) approximation to this posterior, which allows practitioners to assess the overall accuracy of their inferred classes, identify areas with low certainty, and retain practically usable uncertainty quantification.  A review of variational inference for statisticians was written by~\citet{blei_variational_2017}. One of the key strengths of our approach lies in its ability to transfer knowledge from an external region, for which (noisy) labels are available, to a target region, without any labels from the target region being required. Our model can further identify clusters in the target region that are not present in the external labelled data. 

The outline of the paper is as follows. In Section~\ref{sec:statistical_setting}, we introduce our model in detail. Section~\ref{sec:VI} proposes a variational inference method to approximate the posterior. The properties of our model and methodology are illustrated in a numerical simulation study in Section~\ref{sec:numerical_simulations}, where we show the benefit of generalising the Potts model, the ability to detect new clusters, and the resulting changes in posterior uncertainty. In Section~\ref{sec:case_study}, we employ our methodology in a land cover classification problem in North East Scotland. Finally, discussions and further research directions are presented in Section~\ref{sec:discussion}.

It is convenient to fix some notations that will be used throughout the paper.  We write $[k]:=\{1,2,\ldots, k\}$ for the set of the first $k$ positive integers.  The Kullback-Leibler (KL) divergence between two probability distributions $p$ and $q$, where $q$ is absolutely continuous with respect to $p$, is written as $\KL(q||p):=\IE_q[\log(q(x)/p(x))]$.
We write $\mathcal{N}$, $\mathcal{IW}$, $\mathcal{NIW}$, $\mathcal{D}$ and $Cat$ for the normal distribution, inverse wishart, normal inverse wishart, dirichlet and the categorial distribution, respectively. We use Greek letters for model parameters, and Latin letters for variational parameters. $\pi$ denotes the number pi, while the bold symbol $\bm{\pi}$ will be used to represent the vector of mixing proportions $(\pi_1,\ldots,\pi_K)$.

\section{Statistical setting\label{sec:statistical_setting}}
The beginning of this section is dedicated to the standard Potts model, before generalising it in Section~\ref{sec:POTTERS} to allow for non-homogeneous spatial dependences.

\subsection{Baseline Bayesian spatial mixture model}
Consider an image $x\in\IR^{d\times n_1\times n_2}$, where $d$ is the number of observed channels per pixel, and $n_1$ and $n_2$ are the size of the rectangular grid, such that we end up with $n=n_1n_2$ pixels in total.  The restriction to rectangular grids is for convenience, and our approach generalises straightforwardly to other grids. Further suppose that each pixel $x_i\in\IR^d$ falls into one of $K$ classes, which we denote as $z_i\in[K]$.  
As a starting point, we consider the following simplified model, which builds on the framework of~\cite{besag_statistical_1986} and~\cite{celeux2003procedures}, see also the book by~\citet{gelman1995bayesian}.
\begin{align}\begin{split} \label{eq:baseline}
    x_i \vert z_i = k &\sim \mathcal{N}(\mu_k, \Sigma_k) \\
    \mu_k &\sim \mathcal{N}(\eta, \Sigma_k /\kappa ) \\
    \Sigma_k &\sim \mathcal{IW}(\nu, \Psi) \\
    p(z_1, \ldots, z_n \vert \pi, \lambda) & \propto \prod_{i=1}^n \prod_{k=1}^K \pi_k^{\mathbbm{1}(z_i = k)} \cdot \exp\Bigl(\lambda \sum_{j=1}^n  \sum_{h=1}^K W_{ij} \mathbbm{1}(z_i = k, z_j = h, z_i \neq z_k)\Bigr)
    \\
    \pi &\sim \mathcal{D}(\alpha_1,\ldots,\alpha_K).
\end{split}
\end{align}

Here, the interpretation is that each pixel $x_i$ is observed from a normal distribution centred around a \emph{class-mean} $\mu_k$, and where the variance $\Sigma_k$ equally depends on the class.  Both the class-mean and class-variance are given hyperpriors, with the normal distribution and Inverse Wishart distributions being the canonical choices.  The distribution of a pixel's class $z_i$ itself depends further on a fixed prior probability for each class, $\pi_k$, as well as a \emph{Potts parameter} $\lambda$, which controls how likely two neighbouring pixels are clustered differently. $\lambda=0$ indicates to spatial dependence, while $\lambda\rightarrow-\infty$ encourages neighbouring pixels to fall into one class, and $\lambda\rightarrow\infty$ forces neighbouring pixels into different classes. The \emph{weight matrix} $W$ gives high weights to adjacent pixels, and can be chosen in different ways.  One possibility is to let $W_{ij}$ scale inversely with the spatial distance between pixels $i$ and $j$. For computational convenience, we take $W$ to be the adjacency matrix i.\,e. $W_{ij}:=\mathbbm{1}\{i$ is a neighbour of $j\}$, using a queen neighbourhood structure \citep{lloyd2010spatial}. Other choices of $W$ could be e.\,g. $W_{ij}:=\exp(-d(i,j))$, where $d(i,j)$ is the spatial distance between two pixels. 

\subsection{Potts Model for Enhanced Remote Sensing (POTTERS)\label{sec:POTTERS}}
With the availability of externally labelled remote sensing images, it is natural to include any useful information from them in our model, which allows us to generalise the baseline Potts model from the previous section in three ways. First, we introduce a matrix-valued parameter $\Lambda\in\IR^{K\times K}$, which allows class-specific clustering behaviour, taking the role of $\lambda$ in the baseline model, which we can estimate from the external data source.  Second, we use the labelled external data to inform the prior and tuning parameters for the classification problem of the target image $x$.  Third, we allow for the number of clusters in our target image to be unknown, allowing in particular for new classes in the target image that were not present in the external data.

\subsubsection*{Generalised Potts parameter}
To account for the spatial dependence between neighbouring pixels, we define a \emph{generalised Potts model} by enforcing a spatial dependence structure as 
\[ M_\Lambda(z_1, \ldots, z_n) := \dfrac{1}{R(\Lambda)} \exp \Big( \sum_{i=1}^n \sum_{j=1}^n \sum_{k=1}^K \sum_{h = 1}^K \lambda_{kh} W_{ij} \mathbbm{1}\{z_i = k, z_j = h, z_i \neq z_j \} \Big), \]
where $\Lambda = (\lambda_{k,h}) \in \mathbb{R}^{K \times K}$ is the collection of all the interaction parameters in the model, and $W$ contains the information about the neighbourhood structure of the pixels.  Positive values of $\lambda_{k,h}$ indicate that classes k and h are likely to appear next to each other, while negative values indicate they are unlikely to appear next to each other.  The `diagonal' entries are set to be zero for identifiability, i.\,e. $\lambda_{k,k}=0$ for all $k$. 

The conditional distribution of $z$ given $\bm{\pi}$ and $\Lambda$ under the spatial model is then given as 
\[
p(z_1, \ldots, z_n \vert \bm{\pi}, \Lambda)= \dfrac{1}{S(\bm{\pi}, \Lambda)}\prod_{i=1}^n \prod_{k=1}^K \pi_k^{\mathbbm{1}(z_i = k)} \cdot \exp\Bigl(\lambda_{kh} \sum_{j=1}^n \sum_{k=1}^K W_{ij} \mathbbm{1}(z_i = k, z_j = h, z_i \neq z_j)\Bigr),
\]
where $S(\bm{\pi}, \Lambda)$ is a normalisation term depending on both $\bm{\pi}$ and non-fixed $\Lambda$.  We will see later that inference in our full model requires a variational approximation to make it computationally tractable, which is hindered by the computationally prohibitively expensive evaluation of $S(\bm{\pi}, \Lambda)$.  

This generalisation of the Potts model is interesting in its own right, but we focus on the implications for remote sensing here, particularly focusing on land cover classification.  It is well known that land cover classes exhibit different clustering behaviours~\citep{tobler1970computer}, e.\,g. one is unlikely to find mountainous rocks next to saltwater or saltwater marsh, agricultural land often appears in larger clusters than bogs. This phenomenon is also referred to as spatial autocorrelation and spatial interpolation methods such as Kriging~\citep{matheron1963principles}. Therefore, it is reasonable to assume that a class appears more likely next to itself than to any other class;  we thus restrict our admissible values of $\Lambda$ to be negative, i.e. all $\lambda_{k,h}<0$ for $k\neq h$. 

Since the prior $\bm{\pi}$ plays only a minor role in defining the target posterior, and to avoid imposing unnecessary complexity on the model, we consider $\bm{\pi}$ as a fixed parameter instead of sampling from the Dirichlet distribution as in the baseline model~\eqref{eq:baseline}; we instead prioritise the incorporation of spatial correlations through the hyperparameter $\Lambda$. If one is interested in deriving uncertainty quantification for both $\bm{\pi}$ and $\Lambda$ by placing hyperpriors on these quantities, one possibility is to use generalised Bayesian posteriors, see e.\,g.~\citet{laplante_conjugate_2025} for recent work on this. 

\subsubsection*{Informed prior}
The assumption that the class means $\mu_k$ are centred around a joint mean parameter $\eta$, as suggested by the baseline model~\eqref{eq:baseline}, is overly restrictive for our purposes.  We relax this by allowing a different prior for each class mean, $\mu_k  \overset{\mathrm{ind}}{\sim} N(\eta_k, \Sigma_k)$.  This allows incorporating prior information for each cluster component, based on any available cluster-specific knowledge.  We will think of this as an informative prior, where each $\eta_k$ and $\Sigma_k$ incorporates prior knowledge about the cluster-specific distribution, which we will estimate from the external data.  Note that the Bayesian prior-posterior update formula provides a coherent framework to update prior beliefs from available data, and this strategy has been used previously, see e.\,g. the power prior \citep{ibrahim2000power}. 

We now construct an auxiliary model for the external labelled data. To this end, let $\tilde{x} = \{\tilde{x}_{1}, \dots, \tilde{x}_{\tilde{n}}\}$ denote the available external data, and let $(\tilde{z}_1, \ldots, \tilde{z}_{\tilde{n}})$ denote the vector of known cluster allocations for the external data, with $\tilde{z}_i \in [K]$.  We suppose that the external data was generated as
\begin{align*}
\tilde{x}_{i} \vert \tilde{z}_i=k \sim \mathcal{N}(\mu_k, \Sigma_k), \qquad
\mu_k \vert \Sigma_k \sim \mathcal{N}(\eta, \Sigma_k/\kappa), \qquad
\Sigma_k \sim \mathcal{IW}(\nu, \Psi).
\end{align*}

Since the class distributions are independent, we can get the posterior for each of them individually; the posterior of the parameters given the external data is thus given by 
\begin{align*}
p(\mu_1, \ldots, \mu_K, &\Sigma_1, \ldots, \Sigma_K \vert \tilde{x}, \tilde{z}, \eta, \kappa, \nu, \Psi) \\
&\propto  \, \prod_{i\in[n]}p( \tilde{x}_i \vert \mu_{\tilde{z}_i}, \Sigma_{\tilde{z}_i}, \tilde{z}_i)  \cdot \prod_{k\in[K]} \Bigl\{ p(\mu_k \vert \eta, \kappa, \Sigma_k) p( \Sigma_k \vert \nu, \Psi) \Bigr\} \\
&\propto  \prod_{k\in[K]} p\bigl(\mu_k, \Sigma_k \vert \{\tilde{x}_i: \tilde{z}_i=k \}, \eta,\kappa,\nu,\Psi\bigr).
\end{align*}

For each cluster $k$, the parameters $(\mu_k,\Sigma_k)$ have a conjugate Normal-Inverse-Wishart distribution, so its posterior is also a Normal-Inverse-Wishart, which can be written as $\mathcal{NIW}(\tilde{\eta}_k, \tilde{\kappa}_k, \tilde{\nu}_k, \tilde{\Psi}_k)$, where
\begin{align}
\begin{split}
\label{eq:external_data_posterior}
\tilde{\eta}_k &= \frac{\kappa}{\kappa+ \tilde{n}_k} \eta + \frac{\tilde{n}_k}{\kappa+ \tilde{n}_k} \bar{\tilde{x}}_k,\qquad
\tilde{\kappa}_k = \kappa + \tilde{n}_k, \qquad
\tilde{\nu}_k =  \nu + \tilde{n}_k,
\\
\tilde{\Psi}_k &= \Psi + \tilde{n}_k\tilde{S}_k + \frac{\kappa\tilde{n}_k}{\kappa+\tilde{n}_k}(\bar{\tilde{x}}_k-\eta)(\bar{\tilde{x}}_k-\eta)^T.
\end{split}
\end{align}

Here, $\tilde{n}_k = \sum_{i:\tilde{z}_i = k}1$ is the number of observations in cluster $k$, $\bar{\tilde{x}}_k = \frac{1}{\tilde{n}_k} \sum_{i:\tilde{z}_i = k} \tilde{x}_i$ is the cluster-specific average, and $\tilde{S}_k:=\dfrac{1}{\tilde{n}_k}\sum_{i: \tilde{z}_i = k} (\tilde{x}_i - \bar{\tilde{x}}_k)(x_i - \bar{\tilde{x}}_k)^T$ is the sum of squares matrix for observations in cluster $k$.

\subsubsection*{Unknown cluster number}
The informed prior described above aids the classification of pixels that fall into one of the clusters observed in the external data.  There may, however, be pixels that do not fall into any of the external data's clusters, such that we are interested in allowing for ``new'' clusters in our target image. 

We denote $\tilde{K}$ as the number of old clusters and $K'$ as the number of new clusters, so the total number of classes is $K = \tilde{K} + K'$. Since we have no information about unknown clusters, non-informative priors should be applied for cluster means and variances, as in the external data 
\[
(\mu_k, \Sigma_k)_{k \in [K']} \sim \mathcal{NIW}(\eta, \kappa, \nu, \Psi).
\]

The $\Lambda$ parameter from the generalised Potts model, therefore, should be extended to $\Lambda^*$, where
\begin{align}\label{eq:Lambda_def}
\Lambda^* &= \{\lambda^*_{k,h}\}_{k \in [K], h \in [K]}
\begin{cases}
    0  \quad &\text{if} \quad k=h
    \\
    \lambda_{k,h} \quad &\text{if} \quad k \neq h, \quad k \leq \tilde{K} \quad \text{and} \quad h \leq \tilde{K}
    \\
    \lambda \quad &\text{if} \quad k \neq h, \quad k > \tilde{K} \quad\text{or}\quad h > \tilde{K}.
\end{cases}
\end{align}

Since we choose the external data with similar class behaviours as those assumed in the target data, the probability of a pixel belonging to a new cluster should be smaller than the probability of the pixel belonging to a class existing in the external image; otherwise, the model will assign most pixels in the target image to new clusters.  Therefore, $\bm{\pi}$ should be carefully chosen to avoid introducing substantial bias into the model.  We will discuss how to choose the hyperparameters $\Lambda^*_{k,h}$ and $\bm{\pi}$ hyperparameters for each kind of clusters in detail in Section~\ref{sec:hyperparameter_spec}. 

Several studies have explored Bayesian nonparametric methods to allow a non-fixed number of clusters. The Dirichlet process, which was first introduced by \citet{ferguson1973bayesian} and applied to mixture modelling by \citet{escobar1995bayesian}, provides a principled framework for inferring the number of clusters from data. Building on this, \citet{orbanz_nonparametric_2008} combined Dirichlet Process Mixtures with Markov Random Fields to enforce spatial smoothness whilst automatically determining the number of clusters. \citet{da2016bayesian} further extended this by placing a prior on partitions to jointly control the number and size of clusters. Another approach, called the Chinese Restaurant Process, was also proposed for non-fixed clustering, including distance-dependent variants \citep{blei2011distance} and applications to remote sensing imagery \citep{mao2016generalized, shu2016clustering}. However, none of these works incorporates class-specific spatial correlation through a Potts model, leverages informative priors on cluster means and covariances derived from labelled data, accounts for distribution shift between domains, or provides a mechanism for discovering entirely new clusters beyond those observed during training.

\subsubsection*{Potts Model for Enhanced Remote Sensing}
By combining the three generalisations above, we are now in a position to define the Potts Model for Enhanced Remote Sensing (POTTERS) to infer the classes $z$ for an image $x$ using a prior based on external labelled data $(\tilde{x},\tilde{z})$.  The model introduces additional parameters, $\zeta_k$, which we call the \emph{trust} parameters. They allow the user to specify how similar they believe the external and target distributions to be, see our discussion below.  Our full model is given as
\begin{align}\begin{split}
    x_i \vert z_i = k &\sim \mathcal{N}(\mu_k, \Sigma_k)
    \\
    \mu_k \vert \Sigma_k &\sim 
    \begin{cases}
        \mathcal{N}(\eta_k, \Sigma_k / \zeta_k),  &\text{for} \quad 1 \leq k \leq \tilde{K} 
        \\
        \mathcal{N}(\eta, \Sigma_k/\kappa), & \text{for} \quad \tilde{K} +1 \leq k \leq K
    \end{cases}
    \\
    \eta_k \vert \Sigma_k &\sim \mathcal{N}(\tilde{\eta}_k, \Sigma_k / \tilde{\kappa}_k ), \quad \text{for} \quad 1 \leq k \leq \tilde{K} 
    \\ 
    \Sigma_k &\sim 
    \begin{cases}
        \mathcal{IW}\Biggl( \dfrac{\zeta_k \tilde{\nu}_k + d + 1}{\zeta_k + 1}, \dfrac{\zeta_k}{\zeta_k + 1} \tilde{\Psi}_k \Biggr), &\text{for} \quad 1 \leq k \leq \tilde{K} 
        \\
        \mathcal{IW}\bigl(\nu, \Psi \bigr), & \text{for} \quad \tilde{K} +1 \leq k \leq K
    \end{cases} 
    \\
    p(z_1, \ldots, z_n|\bm{\pi},\Lambda^*) &\propto \prod_{i=1}^n \prod_{k=1}^K \pi_k^{\mathbbm{1}(z_i = k)} \cdot \exp \Bigl( \displaystyle\sum_{j \in N(i)} \displaystyle \sum_{h=1}^{K} \lambda^*_{k h} \, \mathbbm{1}\{ z_i = k, z_j = h, z_i \neq z_j \} \Bigr).
    \label{eq:main_model}
\end{split}
\end{align}
Note that the prior hyperparameters for $\mu_k$ and $\Sigma_k$ are implicitly included through the parameters derived from the external data that are fixed $(\tilde{\eta}_k, \tilde{\kappa}_k,\tilde{\nu}_k,\tilde{\Psi}_k)$ and correspond to the parameters of the posterior distribution for the external data defined in Equation~\eqref{eq:external_data_posterior}. The $\zeta_k$ are \emph{trust hyperparameters}, which control, for each class, how much the external data is believed to be following the same class-conditional distributions as the data of interest. A large value for $\zeta_k$ places large trust in the external data, i.\,e the means of the classes in the target image are assumed to be the same as those in the external one; while small values of $\zeta_k$ have the opposite effect, allowing larger deviations of the target means from the observed class means in the external data.  Furthermore, the prior mean of $\Sigma_k$ does not depend on $\zeta_k$, while the variance of $\Sigma_k$ decreases with larger values of $\zeta_k$, again corresponding to larger trust in the external data. We then adjust the prior parameters for $\Sigma_k$ as 
\begin{align}\begin{split}
\label{eq:adjusted_nu_psi_tilde}
    \tilde{\nu}^{\text{adj}}_k & = \dfrac{\zeta_k \tilde{\nu}_k + d + 1}{\zeta_k + 1}
    \\
    \tilde{\Psi}^{\text{adj}}_k &= \dfrac{\zeta_k}{\zeta_k + 1} \tilde{\Psi}_k.
\end{split}\end{align}
In Section~\ref{sec:hyperparameter_spec}, we propose a practical procedure on how to establish the magnitude of the trust parameters $\zeta_k$. 

To simplify notation, we can combine the second and third distributions for existing clusters in~\eqref{eq:main_model} to 
\begin{align}\begin{split}
    \mu_k \vert \Sigma_k &\sim \mathcal{N}\bigl(\tilde{\eta}_k, (\tilde{\kappa}_k^{-1} + \zeta_k^{-1})\Sigma_k  \bigr), \quad \text{for} \quad 1 \leq k \leq \tilde{K},
\end{split}
\end{align}
and define $\tilde{\beta}_k := (\tilde{\kappa}_k^{-1} + \zeta_k^{-1})^{-1}$. Incorporating these transformed variables, our model can equivalently be written as
\begin{align}\begin{split}
    x_i \vert z_i = k &\sim \mathcal{N}(\mu_k, \Sigma_k)
    \\
    \mu_k \vert \Sigma_k &\sim 
    \begin{cases}
        \mathcal{N}(\tilde{\eta}_k, \Sigma_k / \tilde{\beta}_k),  &\text{for} \quad 1 \leq k \leq \tilde{K} 
        \\
        \mathcal{N}(\eta, \Sigma_k/\kappa), & \text{for} \quad \tilde{K} +1 \leq k \leq K
    \end{cases}
    \\ 
    \Sigma_k &\sim 
    \begin{cases}
        \mathcal{IW} \bigl(\tilde{\nu}^{\text{adj}}_k,\tilde{\Psi}^{\text{adj}}_k \bigr), &\text{for} \quad 1 \leq k \leq \tilde{K} 
        \\
        \mathcal{IW}\bigl(\nu, \Psi \bigr), & \text{for} \quad \tilde{K} +1 \leq k \leq K
    \end{cases} 
    \\
    p(z_1, \ldots, z_n) &\propto \prod_{i=1}^n \prod_{k=1}^K \pi_k^{\mathbbm{1}(z_i = k)} \cdot \exp \Bigl( \displaystyle\sum_{j \in N(i)} \displaystyle \sum_{h=1}^{K} \lambda^*_{k h} \, \mathbbm{1}\{ z_i = k, z_j = h, z_i \neq z_j \} \Bigr).
    \label{eq:final_model}
\end{split}
\end{align}

The posterior distribution of interest can thus be written as 
\begin{align}\begin{split}
&p(z, \mu_1,\ldots,\mu_K, \Sigma_1,\ldots,\Sigma_K \mid x) \propto \prod_{i\in[n]}\exp\bigl(-(x_i-\mu_{z_i})^T(\Sigma_{z_i})^{-1}(x_i-\mu_{z_i})\bigr)
\\
&\hspace{8mm} \cdot \prod_{k\in[\tilde{K}]}\exp\bigl(-(\mu_k-\tilde{\eta}_k)^T\tilde{\beta}_k(\Sigma_{k})^{-1}(\mu_k-\tilde{\eta}_k)\bigr) \cdot \prod_{k\in[K']}\exp\bigl(-(\mu_k-\eta)^T\kappa(\Sigma_{k})^{-1}(\mu_k-\eta) \bigr) 
\\
&\hspace{8mm}\cdot \prod_{k\in[\tilde{K}]}|\Sigma_k|^{-(\tilde{\nu}^{\text{adj}}_k+d+1)}\exp\bigl(-\mathrm{tr}(\tilde{\Psi}^{\text{adj}}_k\Sigma_k^{-1})/2\bigr) \cdot \prod_{k\in[K']}|\Sigma_k|^{-(\nu+d+1)}\exp\bigl(-\mathrm{tr}(\Psi \Sigma_k^{-1})/2\bigr) 
\\
&\hspace{8mm} \cdot \prod_{i \in [n]} \prod_{k\in[K]} \pi_k^{\mathbbm{1}(z_i = k)} \cdot \exp \Bigl( \displaystyle\sum_{j \in N(i)} \displaystyle \sum_{h\in[K]} \lambda^*_{k h} \, \mathbbm{1}\{ z_i = k, z_j = h, z_i \neq z_j \} \Bigr) . 
\label{eq:posterior}
\end{split}
\end{align} 

For the interested reader, we provide both a conditional formulation and a matrix formulation of our model in the supplementary material in ~\ref{sec:conditional_formulation_potters} and~\ref{sec:matrix_formulation_potters}, respectively. 

\subsection{Hyperparamater specification\label{sec:hyperparameter_spec}}
We fix the hyperparameters $\eta$, $\kappa$, $\nu$, and $\Psi$ in our model based on the user's domain knowledge; see also \cite{fraley2007bayesian} for default recommendations on these parameters.
The parameters $\tilde\eta$, $\tilde\kappa$, $\tilde\nu$, and $\tilde\Psi$ are computed as presented in Equation~\eqref{eq:external_data_posterior}, and the transformed parameters $\tilde\nu^{\text{adj}}$ and $\tilde\Psi^{\text{adj}}$ are subsequently calculated using~\eqref{eq:adjusted_nu_psi_tilde}. In the remainder of this subsection, we describe our procedure of estimating $\Lambda^*$ and $\bm{\pi}$ in a data-driven way from the external data and discuss how a non-expert can choose appropriate trust parameters $\zeta_k$.

\subsubsection*{Estimation of $\Lambda^*$}
Recall that $\Lambda = (\lambda_{1,1},\lambda_{1,2},\ldots \lambda_{\tilde{K},\tilde{K}})$ collects all generalised Potts parameters, and that $\lambda$ is a single parameter describing an average interaction strength which is used for new classes. We separately estimate $\Lambda$ and $\lambda$ from the external data.

We would like to estimate $\Lambda$ in the generalised Potts model through the external data by solving the corresponding optimisation problem 
\[
\mathrm{max}_{\Lambda} \Bigg[\dfrac{1}{R(\Lambda)} \exp \Big( \sum_{i=1}^{\tilde{n}} \sum_{j=1}^{\tilde{n}} \sum_{h=1}^{\tilde{K}} \lambda_{\tilde{z}_i h} W_{ij} \mathbbm{1}\{\tilde{z}_j = h,\tilde{z}_i \neq \tilde{z}_j \} \Big) \Bigg].
\]
Notice that computing the normalisation constant $R(\Lambda)$ is computationally intractable, as we would need to sum over all possible label permutations.
Recent works have proposed better solutions for solving such intractable likelihood problems. \cite{matsubara_generalized_2024} introduced the Discrete Fisher Divergence (DFD) based on the score rule to calculate the likelihood without concerning the intractable normalisation term. However, this approach only works with binary class cases, where the class order is irrelevant. Log-Ratio Matching Divergence was recently introduced in \cite{laplante_conjugate_2025}, but retains the same problem as the DFD. The Maximum Pseudo-likelihood Estimation (MPLE) method \citep{zhou_bayesian_2009, li_stan_z_markov_2009} is one of the most widely used approaches as it is simple and computationally efficient when ignoring the normalisation term. Instead of optimising the full likelihood, MPLE only maximises the pseudo-likelihood that is defined in our case as
\[
\tilde{L}_\Lambda(z) = \sum_{i=1}^{\tilde{n}} \log\Biggl( \dfrac{ \exp \Big( \displaystyle \sum_{j \in N(i)}  \displaystyle \sum_{h=1}^{\tilde{K}} \lambda_{\tilde{z}_i h} \mathbbm{1}\{ \tilde{z}_j = h, \tilde{z}_i \neq \tilde{z}_j \} \Bigr) }{ \displaystyle \sum_{k'=1}^{\tilde{K}} \exp \Big( \sum_{j \in N(i)}  \displaystyle \sum_{h=1}^{\tilde{K}} \lambda_{k'h} \mathbbm{1}\{ \tilde{z}_i = k', \tilde{z}_j = h, \tilde{z}_i \neq \tilde{z}_j \} \Bigr) } \Biggr).
\]
The optimal $\widehat{\Lambda}$ is obtained by solving the optimisation problem
\begin{equation}
    \widehat{\Lambda} = \argmax_{\Lambda} \big[\tilde{L}_\Lambda(z) \big],
\label{eq:optimisation_problem}
\end{equation} 
where $\widehat\Lambda$ has the form as in Equation (\ref{eq:Lambda_def}). However, this optimisation problem is ill-defined, which is shown in the following theorem.
\begin{theorem}
    There does not exist $\widehat{\Lambda} = \{\widehat{\lambda}_{kh}\}_{k \in [K], h \in [K]}$ with $\widehat{\lambda}_{k,h} \in \mathbb{R} $ satisfying
    \[\widehat{\Lambda} = \argmax_{\Lambda^*} \big[\tilde{L}_\Lambda(z) \big]. \]
    \label{thm: thm1}
\end{theorem}
The proof for Theorem \ref{thm: thm1} can be found in Appendix \ref{sec:thm_proof}. Therefore, we need a regularisation term to ensure well-posedness of the optimisation problem. Let us define 
\[ g(\Lambda) = \tilde{L}_\Lambda(z) - \rho \, \|\Lambda\|_F^2, \]
where $\|\Lambda\|_F := \sqrt{\sum_{k=1}^{K} \sum_{h=1}^{K} |\lambda_{kh}|^2}$, and $\rho >0$ is the regularisation coefficient. A large $\rho$ will shrink most of the $\lambda_{k,h}$ to 0, while a small $\rho$ allows larger values of $\lambda_{k,h}$. 
Now, $\widehat\Lambda$ is estimated as $\widehat{\lambda}_{kh}:=\argmax_{\Lambda:\mathrm{diag}(\Lambda) = 0,\lambda_{kh} \le 0 } g(\Lambda)$ for existing clusters, and $\hat\lambda:=\argmax_{\lambda\leq0} g(\lambda\cdot(1-I))$ for new clusters. 

Finally, we set
\begin{align*}
\widehat\Lambda^* = \{\widehat \lambda^*_{kh}\}_{k \in [K], h \in [K]} =  
\begin{cases}
    0  \quad &\text{if} \quad k=h
    \\
    \hat\lambda_{kh} \quad &\text{if} \quad k \neq h, \quad k \leq \tilde{K} \quad \text{and} \quad h \leq \tilde{K}
    \\
    \hat\lambda \quad &\text{otherwise}.
\end{cases}
\end{align*}
In Sections~\ref{sec:numerical_simulations} and~\ref{sec:case_study}, we will discuss how to choose the regularisation parameter $\rho$ in specific examples.

\subsubsection*{Choice of $\bm{\pi}$}
The vector of prior class probabilities $\bm{\pi}$ has to be chosen carefully to prevent model bias. Assigning excessive mass to existing clusters will limit the model's flexibility to find new ones, while over-weighting new clusters will tell the model to prefer new clusters and force all pixels into them. Therefore, we define $\bm{\pi} = (\pi_1, \ldots, \pi_K)$ as
\begin{align*}
    \pi_k &= 
    \begin{cases}
        \epsilon \cdot \dfrac{\tilde{n}_k}{n} + (1 - \epsilon)\cdot\dfrac{1}{K} \, & \text{for} \quad 1 \leq k \leq \tilde{K}  
        \\
        (1-\epsilon)\cdot \dfrac{1}{K} \, &\text{for} \quad \tilde{K} +1 \leq k \leq K,
    \end{cases}
\end{align*}
where $\epsilon$ is a chosen parameter governing the balance between existing and new clusters, and $\tilde{n}_k$ is the number of pixel observations belonging to cluster $k$ in the external data. In this way, we can control how strongly we believe that new clusters are likely to exist.  In Sections~\ref{sec:numerical_simulations} and~\ref{sec:case_study}, we will specify the choice of $\epsilon$ for simulated and real-world data problems.

\subsubsection*{Choice of $\zeta_k$}
The trust parameter $\zeta_k$ reflects how much trust we place in the labelled external data, that is, how similar we expect the class distributions between the external dataset and our target data to be. As this trust is usually difficult to assess in practice, we here propose an intuitive procedure to choose the $\zeta_k$. 

We first run our POTTERS model with three different values for $\zeta_k$, where, respectively, $\zeta_k \in \{0.01, 1, 100\}$ for all $k$.  Here, $\zeta_k = 0.01$ reflects little trust in the external data, whereas $\zeta_k = 100$ indicates strong confidence that two distributions are similar, while $\zeta_k = 1$ represents a neutral confidence level.  Based on these three results, we identify classes that appear to be reliable across all runs by comparing them with satellite imagery and domain knowledge, and we set $\zeta_k = 100$ for such classes $k$. For classes that do not align with what we believe to be the ground truth, we set $\zeta_k = 0.01$, while for classes about which we have limited confidence, we set $\zeta_k = 1$.  With these chosen $\zeta_k$, we run the algorithm again and adjust the $\zeta_k$ for each class by multiplying or dividing the current choice for each $\zeta_k$ by 10, if needed, based on our assessment of the accuracy of the individual class predictions. Note that we could repeat this adjustment a few times, at each step comparing the satellite images and any observed changes of cluster assignments in the model output. The details for this process can be found as the pseudo-code in Algorithm~\ref{alg:zeta_alg} in Appendix~\ref{sec:zeta_alg_appendix}.

\FloatBarrier
\section{Variational methodology\label{sec:VI}}
The posterior defined in~\eqref{eq:posterior} will generally be intractable for most data sizes and computers.  Compared to Monte Carlo Markov Chain (MCMC) methods, Variational Inference (VI) \citep{bernardo2003variational,nasios_variational_2006,fox_tutorial_2012,blei_variational_2017} is more computationally efficient, especially when dealing with high-dimensional data. 

The goal of VI is to find an optimal approximation $q(z, \mu, \Sigma)$ to the true posterior $p(z,\mu,\Sigma|x)$ by minimising the KL divergence, i.\,e.
\begin{align}
    q^\star(z,\mu, \Sigma) &= \argmin_q \KL\bigl(q(z,\mu,\Sigma) || p(z,\mu,\Sigma|x \bigr) \notag
    := \argmin_q \mathbb{E}_q\Big[\log\dfrac{q(z,\mu,\Sigma)}{p(z,\mu,\Sigma|x)}\Big].
    \label{eq:variational optimisation}
\end{align} 

The KL divergence can be transformed as 
\begin{align*}
    \KL(q(z)|| p(z|x)) &= \mathbb{E}_q[\log q(z,\mu,\Sigma)] - \mathbb{E}_q[\log p(z,\mu,\Sigma|x)] \\
    &=\mathbb{E}_q[\log q(z,\mu,\Sigma)] - \mathbb{E}_q[\log p(z,\mu,\Sigma, x)] + \log p(x).\\
\end{align*}
We notice that the KL divergence is intractable as it requires computing the logarithm of the evidence $\log p(x)$. If we add another constant to the KL, then the optimisation objective is not affected, which leads to the definition of the evidence lower bound (ELBO) as 
\begin{equation}
    \ELBO(q):= \mathbb{E}_q[\log p(z,\mu, \Sigma, x)] - \mathbb{E}_q[\log q(z, \mu, \Sigma)],
    \label{eq:general elbo}
\end{equation}
and the KL divergence can then be written as 
\begin{align*}
    \KL(q(z,\mu,\Sigma)|| p(z, \mu, \Sigma|x)) = -\ELBO(q) + \log p(x).
\end{align*}
Hence, minimising the KL divergence $\KL(q||p)$ with respect to $q$ is equivalent to maximising the $\ELBO(q)$. 

It remains to choose an appropriate variational family.  The Mean-field Variational approximation~\citep{consonni_mean-field_2007} assumes that the variational family can be factorised into independent variational factors
\begin{align*}
    q(z,\mu_k,\Sigma) = \prod_{i = 1}^N q(z_i) \cdot \prod_{k=1}^K q(\mu_k,\Sigma_k),
\end{align*}
which we choose for its computational advantages. 

The ELBO for this variational problem can be explicitly calculated, see Section~\ref{sec:ELBO} for details. We adopt the Coordinate Ascent Mean-Field Variational Inference (CAVI) algorithm \citep{bishop2006pattern} to solve the optimisation problem. The idea is to update each one of the variational factors at a time while keeping the others fixed, which requires calculating the conditional expectation for each of the variational parameters while the other parameters are fixed. 

The variational distributions can be explicitly written in closed form as shown in Appendix~\ref{sec:appendix variational distribution}, which are 
\begin{align*}
    q(z_i) = Cat(c_{i}), \quad
    q(\mu_k,\Sigma_k) = \mathcal{NIW}(m_k,b_k,v_k,C_k).
\end{align*}
The respective optimal variational parameters are given by
\begin{align}\begin{split}
    c_{ik} &= \dfrac{c'_{ik}}{\displaystyle \sum_{h \in [K]} c'_{ih}}
    \\
    b_k &= 
    \begin{cases}
        \tilde{\beta}_k + N_k &\text{for} \quad 1\leq k \leq \tilde{K} 
        \\
        \kappa + N_k &\text{for} \quad \tilde{K} +1 \leq k \leq K
    \end{cases}
    \\
    m_k &= 
    \begin{cases}
        \dfrac{\tilde{\beta}_k \tilde{\eta}_k + N_k \bar{x}_k}{N_k+ \tilde{\beta}_k} &\text{for} \quad 1\leq k \leq \tilde{K}  
        \\
        \\
        \dfrac{\kappa \eta + N_k \bar{x}_k}{N_k + \kappa} &\text{for} \quad \tilde{K} +1 \leq k \leq K 
    \end{cases}
    \\
    v_k &= 
    \begin{cases}
        \tilde{\nu}^{\text{adj}}_k + N_k &\text{for} \quad 1\leq k \leq \tilde{K} 
        \\
        \nu + N_k &\text{for} \quad \tilde{K} +1 \leq k \leq K
    \end{cases}
    \\
    C_k &= 
    \begin{cases}
        \tilde{\Psi}^{\text{adj}}_k + N_k S_k + \dfrac{N_k \tilde{\beta}_k}{N_k+ \tilde{\beta}_k}  (\bar{x}_k - \tilde{\eta}^{\text{adj}}_k)(\bar{x}_k - \tilde{\eta}^{\text{adj}}_k)^T &\text{for} \quad 1\leq k \leq \tilde{K} 
        \\
        \\
        \Psi + N_k S_k + \dfrac{N_k \kappa}{N_k + \kappa}  (\bar{x}_k - \eta)(\bar{x}_k - \eta)^T &\text{for} \quad \tilde{K} +1 \leq k \leq K,
    \end{cases}
    \label{eq:update_variational_parameters}
\end{split}\end{align}

where $N_k, \bar{x}_k, S_k$ are calculated as
\begin{align}\begin{split}
    N_k = \sum_i c_{ik}, \quad \bar{x}_k &= \dfrac{1}{N_k}\sum_i c_{ik}x_i, \quad S_k = \dfrac{1}{N_k} \sum_i c_{ik}(x_i - \bar{x}_k)(x_i - \bar{x}_k)^T
    \label{eq: common terms}
\end{split}\end{align}

and
\begin{align}
\label{eq:ln_c_prime}
    \ln c'_{ik} = -\frac{1}{2} \mathbb{E}(\ln|\Sigma_k|) -\frac{1}{2} \mathbb{E}[(x_i -\mu_k)^T\Sigma_k^{-1}(x_i - \mu_k)] - \frac{d}{2}\ln 2\pi + \ln \pi_k + \sum_{j \in N(i)} \sum_{h=1}^K \lambda^*_{kh} c_{jh}
\end{align}
with
\begin{align}\begin{split}
    \mathbb{E}[\ln |\Sigma_k|] &=  -d\ln 2 + \ln |C_k| -\sum_{i=1}^d \psi \Bigr( \dfrac{v_k + 1 - i}{2} \Bigl)
    \\
    \mathbb{E}[(x_i-\mu_k)^T\Sigma_k^{-1}(x_i- \mu_k)]& = v_k(x_i - m_k)^T C_k^{-1} (x_i - m_k) + db_k^{-1}.
\end{split}\end{align}

%
%

\begin{algorithm}[H]
\begin{minipage}{\linewidth}
\small
\caption{CAVI for the POTTERS Model\label{alg:potters}}
\begin{algorithmic}[1]
\setlength{\itemsep}{-1pt}

\State \textbf{Inputs:} 
\State \quad Data \(X \in \mathbb{R}^{N \times d}\).
\State \quad \(\tilde{K}\) existing clusters, \( K' \) new clusters, \(K = \tilde{K} + K'\) total number of clusters.
\State \quad Convergence threshold \(\tau\).
\State \quad Fixed prior parameters \((\pi_k, \zeta_k, \kappa, \eta, \nu, \Psi)\).
\State \quad Pre-estimated parameters \((\tilde{\kappa}_k, \tilde{\eta}_k,  \tilde{\nu}_k,\tilde{\Psi}_k,\widehat{\Lambda}^*)\). 

\Statex
\State \textbf{Initialisations:} 
\State \quad \( c^{(0)}, m_k^{(0)}, b_k^{(0)}, v_k^{(0)}, C_k^{(0)} \) are chosen as in \ref{sec:alg_init}

\State \quad \( ELBO^{(0)} \gets -\infty \)
\State \quad \( t \gets 0 \)
\Statex

\Repeat
    \State \( t \gets t + 1 \)    
    \For{ \( i \in [N] \) and \( k \in [K] \)}
        \State Calculate ${c'}_{ik}^{(t)}$ as in \eqref{eq:ln_c_prime}
    \EndFor
    \For{ \( i \in [N] \) and \( k \in [K] \)}
        \State \( c_{ik}^{(t)} \gets {c'}_{ik}^{(t)} / \displaystyle\sum_{h\in[K]} {c'}_{ih}^{(t)} \)
    \EndFor
    \For{\( k \in [K] \)}
        \State Update \( N_k^{(t)}, \bar{x}_k^{(t)}, S_k^{(t)} \) using (\ref{eq: common terms}) with $c^{(t)}$.
        \State Update variational parameters $b_k^{(t)}, m_k^{(t)}, v_k^{(t)}, C_k^{(t)}$ as in (\ref{eq:update_variational_parameters})
    \EndFor
    \State Compute \(\ELBO^{(t)} \) as in (\ref{eq:ELBO}) using the updated variational parameters.
\Until{ \( \dfrac{|\ELBO^{(t)} - \ELBO^{(t-1)}|}{ |\ELBO^{(t-1)}|} < \tau \) }

\Statex
\State \textbf{Outputs:} Optimal variational parameters \(c^{(t)}, m_k^{(t)}, b_k^{(t)}, v_k^{(t)}, C_k^{(t)} \).

\end{algorithmic}
\end{minipage}
\end{algorithm}

The detailed calculations for deriving the variational distributions and the ELBO can be found in Sections~\ref{sec:appendix variational distribution} and~\ref{sec:ELBO} in the Appendix, respectively.  The algorithm is initialised as specified in Appendix~\ref{sec:alg_init} to ensure a warm start, and converges when the relative change in the ELBO falls under a prespecified threshold $\tau$. Pseudocode for the full procedure is detailed in Algorithm~\ref{alg:potters}.

\section{Numerical Simulation Study\label{sec:numerical_simulations}}

In this section, we will conduct a numerical simulation study to illustrate the benefits of our model when compared to the baseline model.  To this end, we simulate both an external (labelled) data set and a target data set.  We will show that including the generalised Potts model substantially improves performance, and that our model indeed detects new clusters present in the target data set. The code is available at \url{https://github.com/bknguyen11/POTTERS/tree/main/simulation_study}.

\begin{figure}[htp]
    \centering
    \begin{subfigure}{\textwidth}
        \centering
        \includegraphics[width=0.8\linewidth]{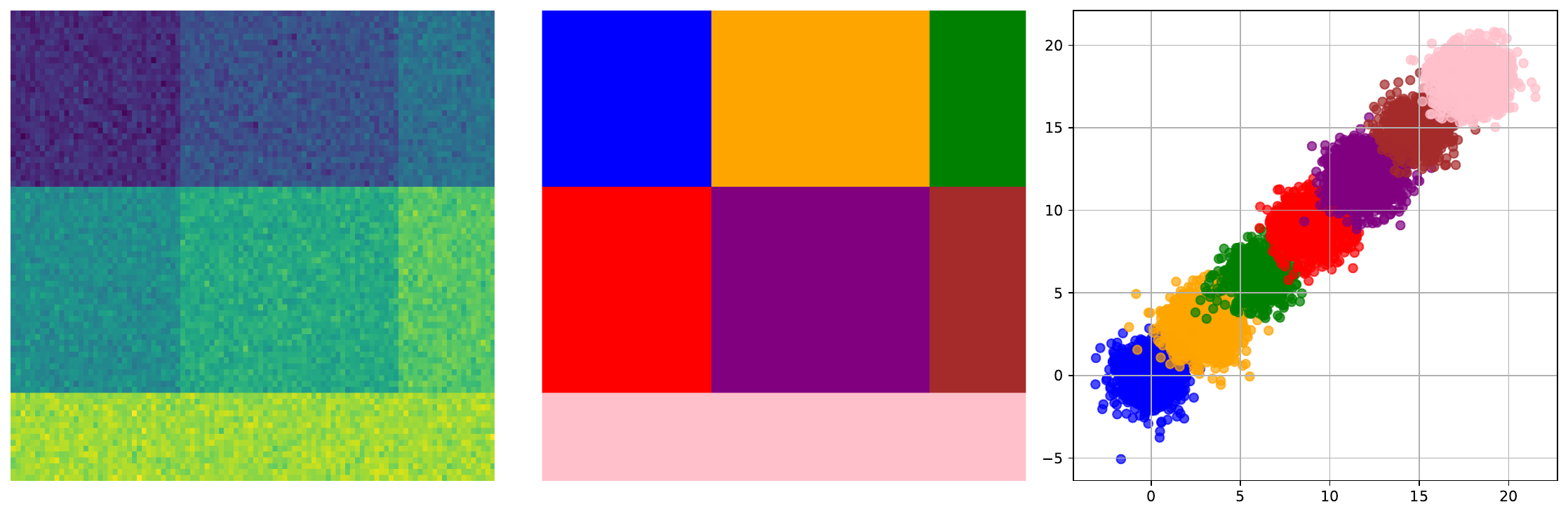}
    \label{fig:sim_pretrain_data}
    \caption{Simulated external data set}
    \end{subfigure}
    \hfill
    \begin{subfigure}{\textwidth}
        \centering
        \includegraphics[width=0.8\linewidth]{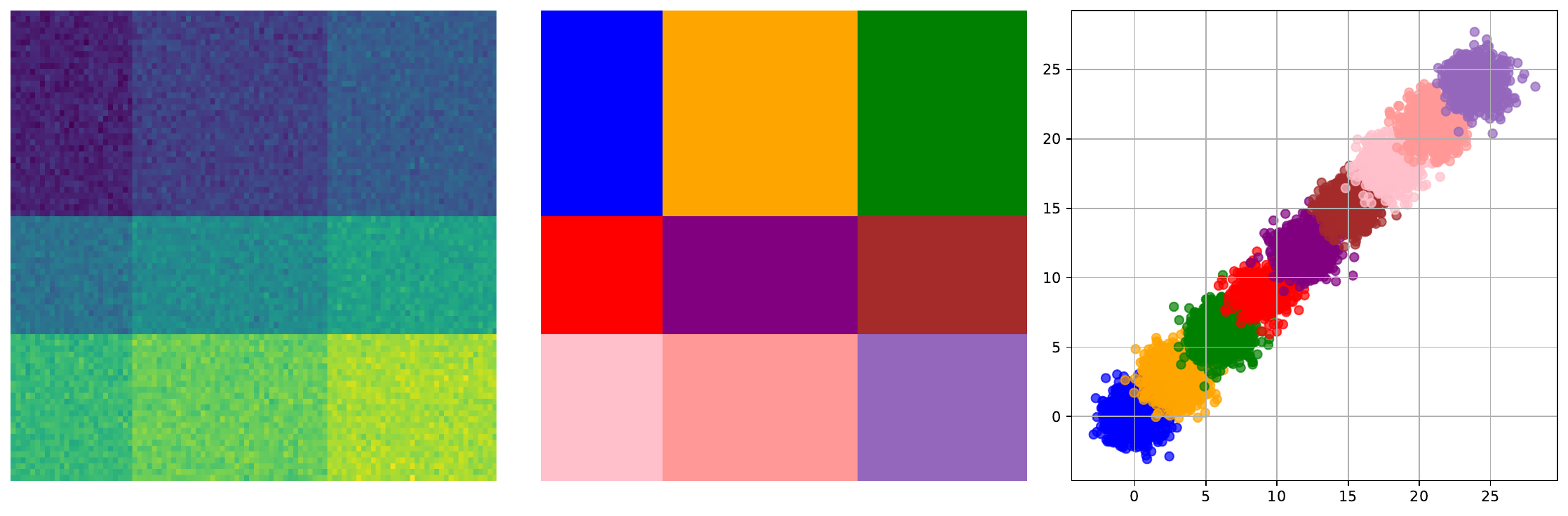}
    \label{fig:sim_train_data}
    \caption{Simulated target data set}
    \end{subfigure}
    \caption{Simulated data sets. For each dataset, the left panel shows a heatmap of the simulated data, mimicking satellite imagery in real-world applications. The middle panel displays the label map, where different colours indicate distinct clusters. The right panel presents a scatter plot illustrating the spatial distribution of the clusters. Each color represents a distinct class. Note that, compared with the external dataset, our target dataset includes two additional classes.}  
    \label{fig:sim_data}
\end{figure}

We first generate labels $\tilde{y}\in \{1,\ldots, 7\}^{100\times 80}$ for the simulated external data set, and labels $y\in\{1,\ldots,9\}^{200\times 100}$ for the simulated target data set.  Next, we simulate the associated images, consisting of two variables per pixel, by sampling each pixel from a Gaussian centered around a label-dependent mean; we denote the simulated external and target images as $\tilde{x}$ and $x$, respectively. See Figure~\ref{fig:sim_data} for an illustration, and see Appendix~\ref{sec:hyperparam_sim_data} for our hyperparameter choices. 

\begin{figure}[htp]
    \centering
    \begin{subfigure}{0.49\textwidth}
        \centering
        \includegraphics[width=0.8\linewidth]{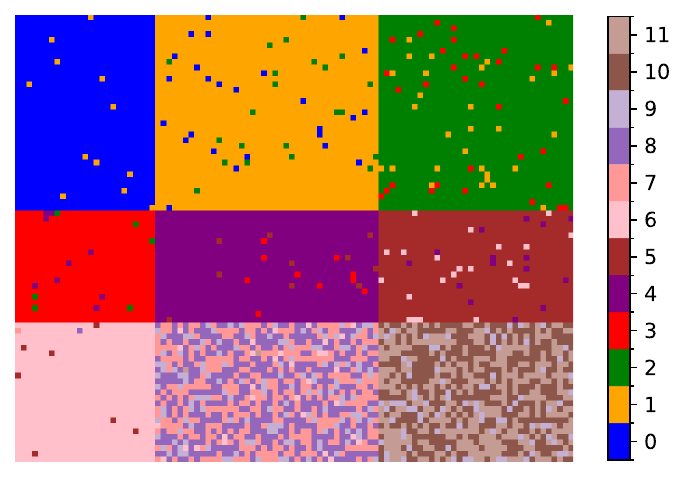}
        \includegraphics[width=0.8\linewidth]{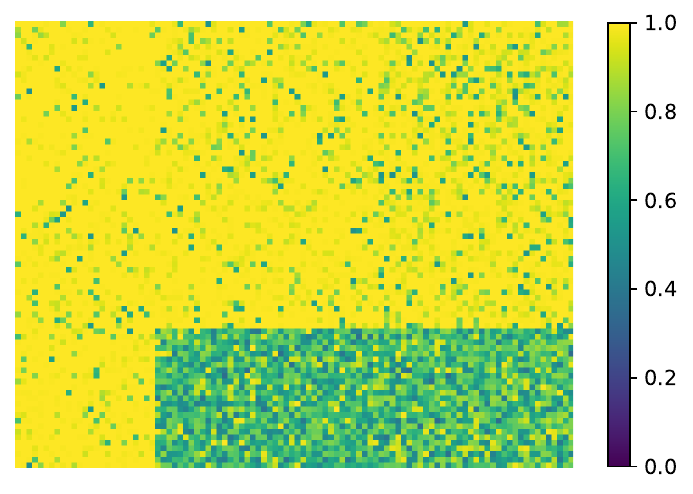}
        \caption{Without Potts model}
        \label{fig:sim_data_result_without_Lambda}
    \end{subfigure}
    \hfill
    \begin{subfigure}{0.49\textwidth}
        \centering
        \includegraphics[width=0.8\linewidth]{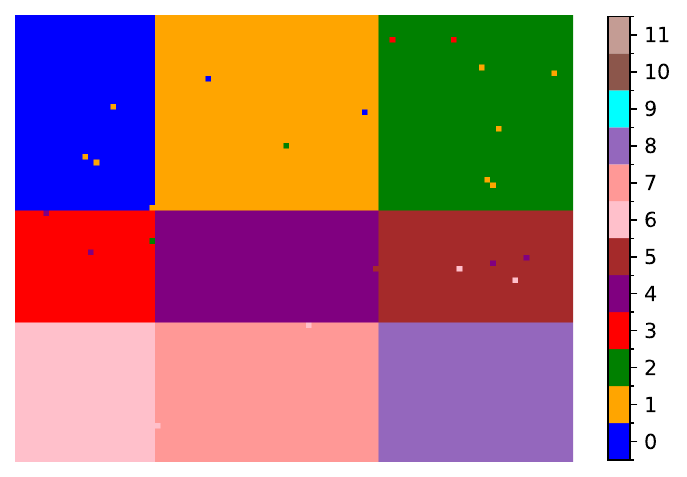}
        \includegraphics[width=0.8\linewidth]{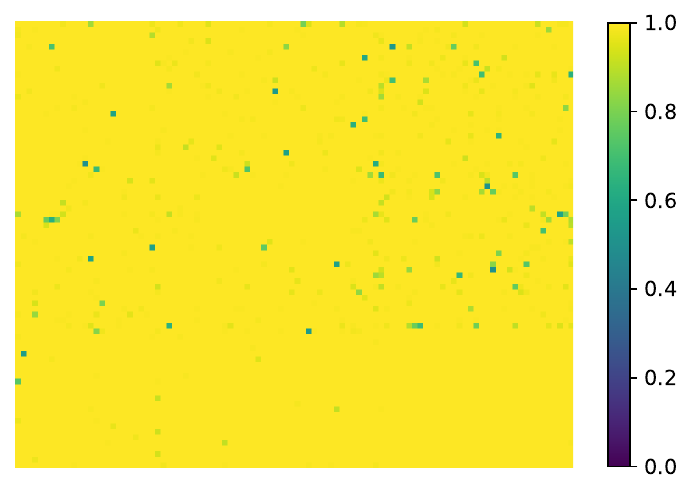}
        \caption{With Potts model}
        \label{fig:sim_data_result_with_Lambda}
    \end{subfigure}
    \caption{The result maps produced by POTTERS in two situations: (a) without the Potts model and (b) with the Potts model. The top row shows the resulting cluster assignment maps. The bottom row displays the corresponding posterior probability maps, indicating the uncertainty of cluster assignments.}
    \label{fig:sim_data_result}
\end{figure}

Our goal is now to estimate the labels $y$ from the external labelled image and the observed target image using our methodology outlined in Section~\ref{sec:POTTERS}.  The hyperparameters, in particular $(\tilde{\eta}_k, \tilde{\kappa}_k, \tilde{\nu}_k, \tilde{\Psi}_k)$, and $\Lambda^*$, are estimated from the external data as described in Section~\ref{sec:hyperparameter_spec}.  We set the trust parameter $\zeta_k = 100$ for every cluster, indicating that we strongly trust the external data. We further set $K' = 5$, i.\,e. to bound the number of new clusters by $5$ (while, of course, we know the true number of new clusters is $2$!), and set $\epsilon = 0.99$, while the regularisation term in the $\Lambda^*$ optimisation problem is set to $\rho = 1$.  We run Algorithm~\ref{alg:potters} twice, once setting $\Lambda^*=0$, corresponding to a model with no spatial information, and once using the estimated $\Lambda^*$, corresponding to our full POTTERS model, to illustrate how including a generalised Potts term aids clustering by incorporating spatial information in the model. In both cases, we terminate the algorithm based on a tolerance of $\tau = 10^{-10}$.  

The results are visualised in Figure~\ref{fig:sim_data_result}.  When the spatial information is ignored, we notice that there are many misclassified pixels appearing in both existing classes and new classes. The algorithm can identify new clusters; however, it cannot well separate these clusters, which explains why in the probability map, these two clusters have significantly lower class probabilities. Looking at the result when we incorporate the Potts model, all the above-mentioned issues are resolved, and the algorithm can find all the clusters properly with high certainty. Although a small number of pixel points are misclassified, these pixels have low assignment probabilities, i.e. larger classification uncertainties.

\section{Case study\label{sec:case_study}}
We now illustrate the performance of our model by classifying land cover in a region in North East Scotland close to Glensaugh. The code for the case study can be found at \url{https://github.com/bknguyen11/POTTERS/tree/main/case_study}.

\subsubsection*{Data}
Our target data image consists of three concatenated satellite images of the target area corresponding to three seasons (Autumn, Spring, and Summer), each with six color channels - Red, Green, Blue, Near-Infrared, Shortwave Infrared 16, Shortwave Infrared 22, as well as the Normalised Difference Water Index~\citep{gao1996ndwi}, giving us a 21-dimensional image.  Winter was omitted due to snow cover preventing the extraction of useful information.
We further make use of an external labelled dataset from the North East of England, which contains similar land cover classes to the Glensaugh region.  The labels are obtained from the Living England Habitat Map (Phase 4)~\citep{kilcoyne2022living}, while the image is a concatenation of three satellite images with the same spectral bands and time period as the target area.  As the external labels are based on satellite images from 2019 and 2020, we use these years for our external imagery data, while the target images are taken from the years 2023 and 2024, coinciding with the time when an ecological study was undertaken, which we will later use for validation. 
To reduce the dimensionality of our imaging data, we conduct a principal component analysis of our images and retain $6$ principal components (compared to $21$ channels in the original image), corresponding to $99\%$ of the variance.  The pre-processing steps to obtain these three images are described in detail in Appendix~\ref{sec:case_study_details}. Figure~\ref{fig:case_study_data} shows satellite images of our target region and of the external region over the spring period (using the Red, Green, and Blue Sentinel channels), and the labels for the external data. All used satellite images are displayed in Figures~\ref{fig:Glensaugh_orginal_data} and~\ref{fig:England_orginal_data} in Appendix~\ref{sec:case_study_details}.

\begin{figure}[htp]
    \begin{minipage}[b]{0.3\textwidth}
        \begin{subfigure}[b]{\textwidth}
        \centering
            \includegraphics[width=0.8\textwidth]{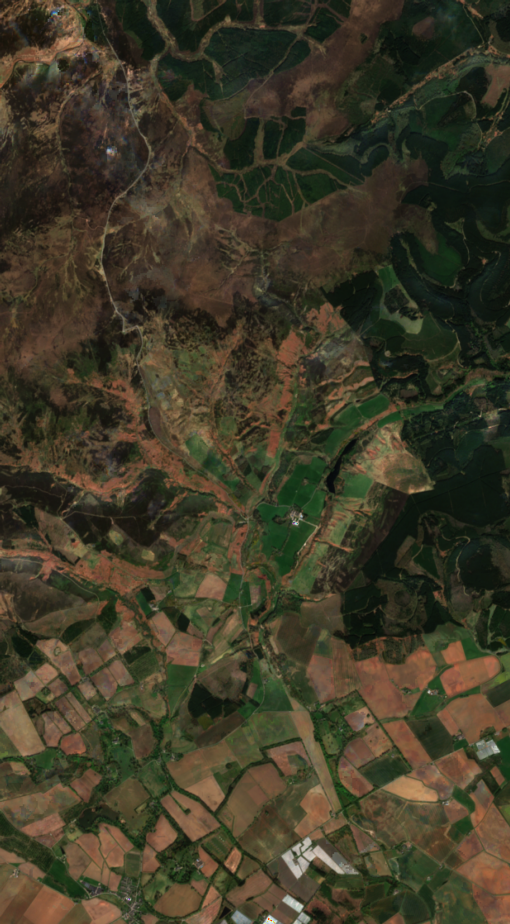}
            \caption{}
            \label{fig:Glensaugh_data_spring}
        \end{subfigure}
    \end{minipage}
    \hspace{0.001\textwidth}
    \begin{minipage}[b]{0.66\textwidth}
    \centering
        \begin{subfigure}[b]{0.43\textwidth}
            \includegraphics[width=0.8\textwidth]{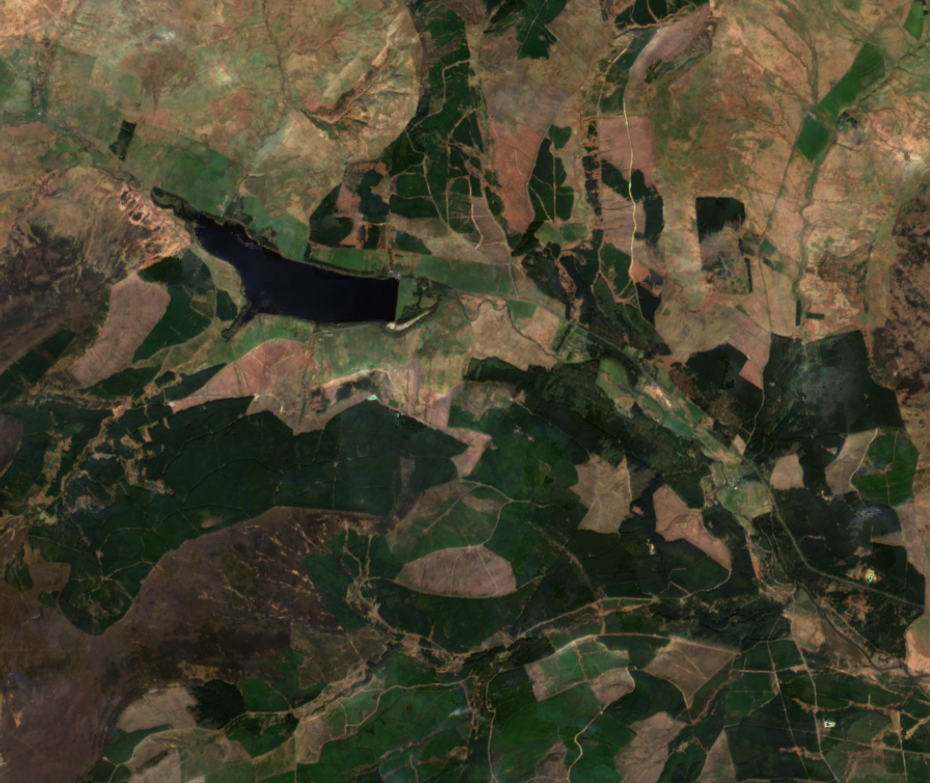}
            \caption{}
            \label{fig:England_data_spring}
        \end{subfigure}
        
        \begin{subfigure}[b]{\textwidth}
        \centering
            \includegraphics[width=0.35\textwidth]{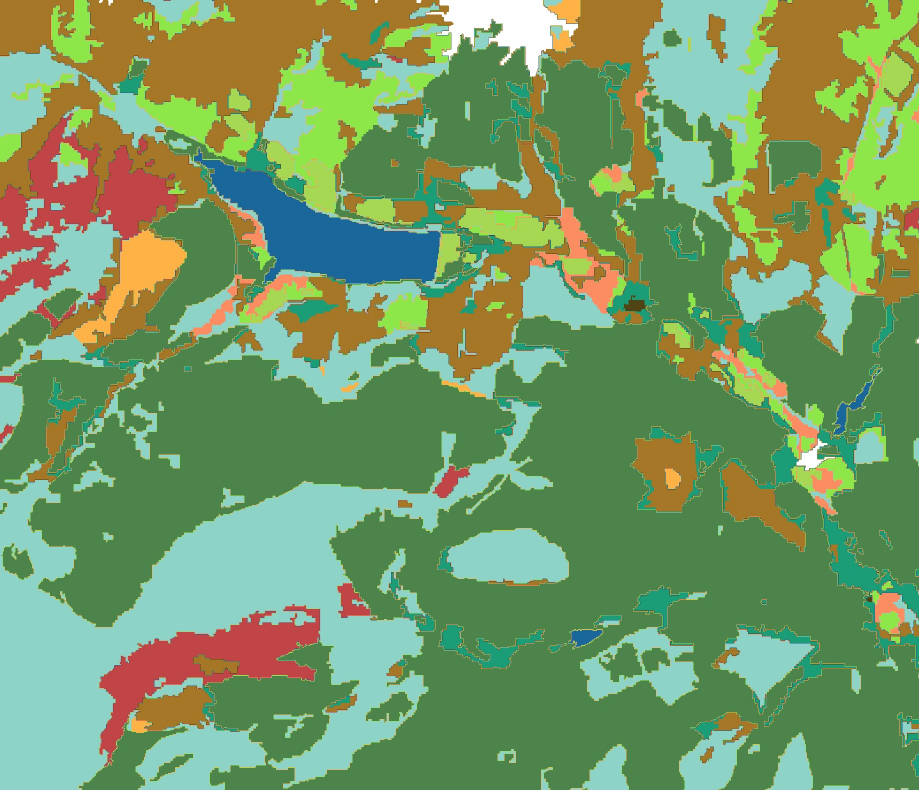}
            \includegraphics[width=0.4\textwidth]{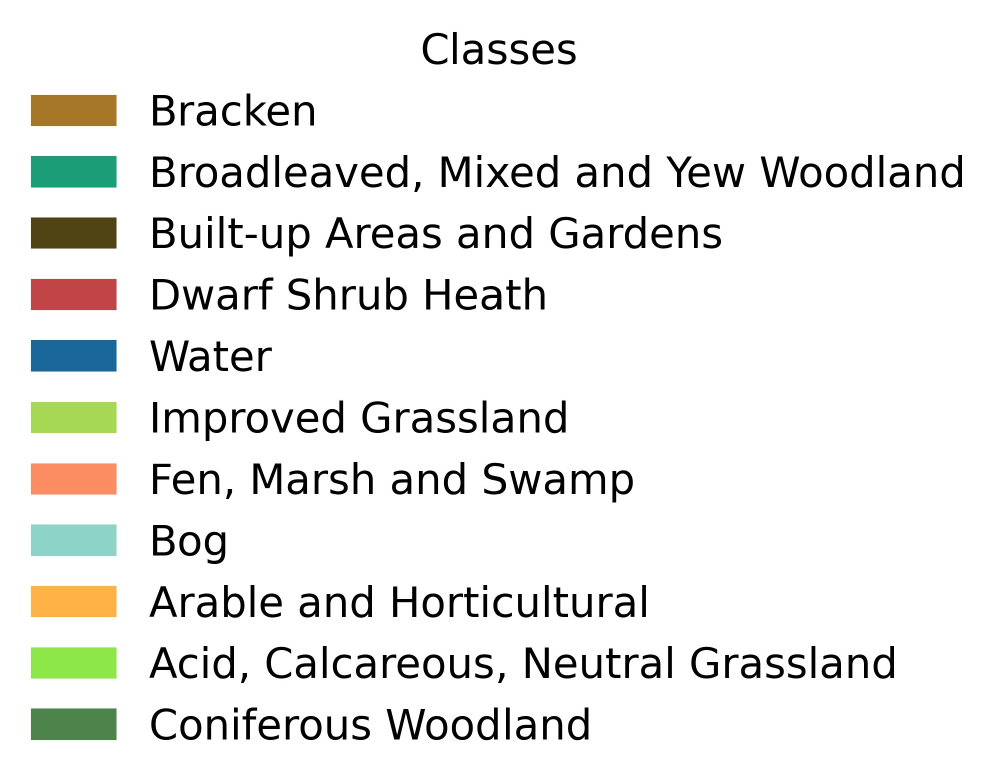}
            \caption{}
            \label{fig:England_labels}
        \end{subfigure}
    \end{minipage} 
    \caption{(a) Glensaugh region satellite image, March–May 2024. (b) England region satellite image, March-May 2020. (c) Labels of the external data, obtained from the Living England Habitat Map.}
    \label{fig:case_study_data}
\end{figure}

\subsubsection*{Hyperparameter choice}
For the hyperparameters in our POTTERS model, the informed prior is based on the principal components of the labelled England image as outlined in Section~\ref{sec:hyperparameter_spec}. The regularisation parameter in the optimisation of $\Lambda$ is chosen as $\rho=1$ to allow the model to capture a broad range of correlations in the data. The trust parameters $\zeta_k$ are chosen as discussed in the same section, see also Algorithm~\ref{alg:zeta_alg}. Furthermore, an ecological expert noted that the Bracken class in the England data is often incorrectly labelled; hence, we set $\zeta = 0.01$ for the Bracken class, indicating little trust in the external data for this class. The weight parameters placed on $\bm{\pi}$ are set to $\epsilon = 0.99$ with the number of new clusters $K'=5$. The convergence tolerance for Algorithm~\ref{alg:potters} is set as $\tau = 10^{-6}$. 

\subsubsection*{Results}
\begin{figure}[htp]
    \centering
    \begin{subfigure}[t]{0.65\linewidth}
        \centering
        \includegraphics[width=\linewidth]{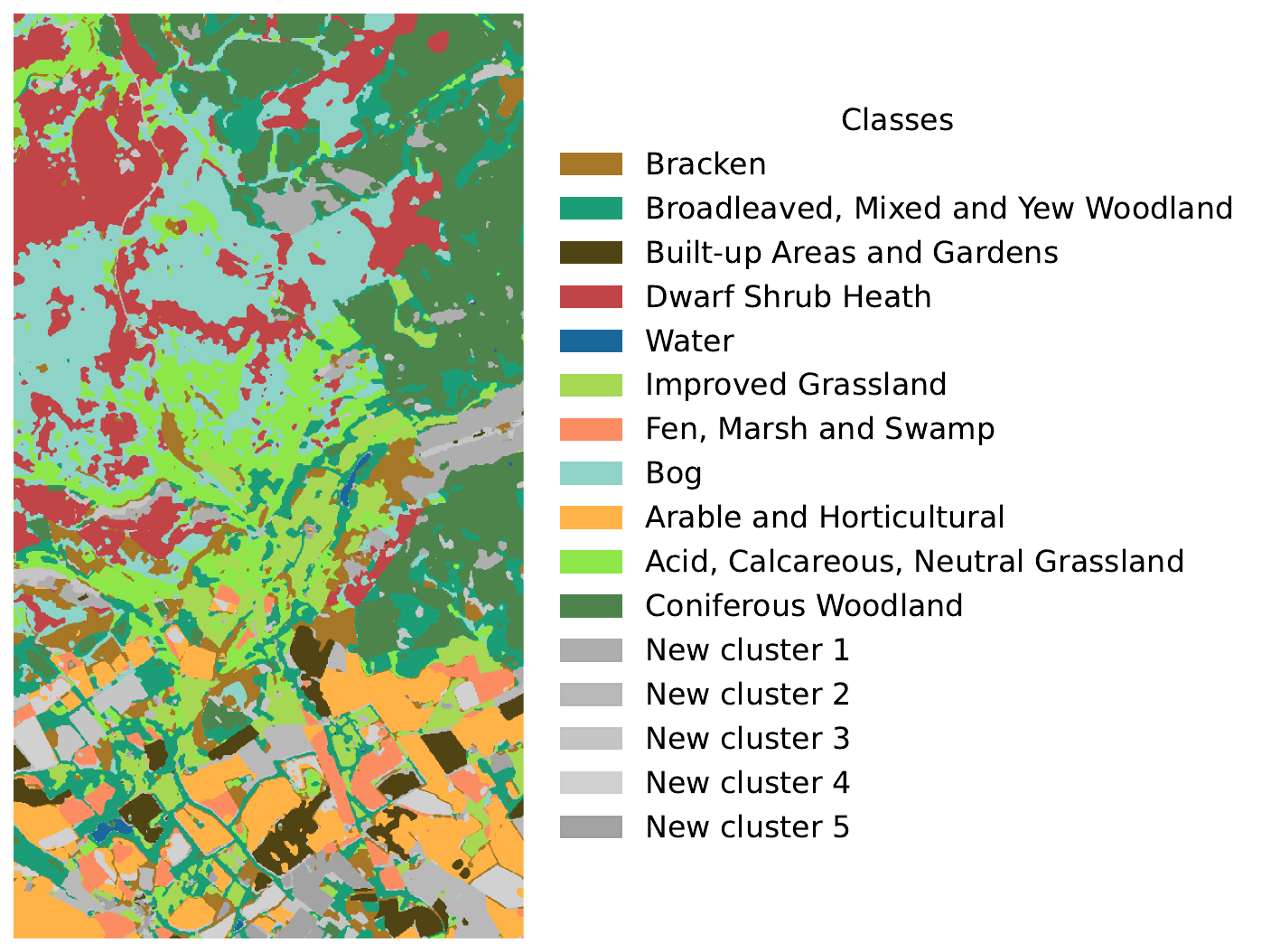}
        \caption{Cluster map \label{fig:Glensaugh_result_class}}
    \end{subfigure}
    \begin{subfigure}[t]{0.34\linewidth}
        \includegraphics[width=\linewidth]{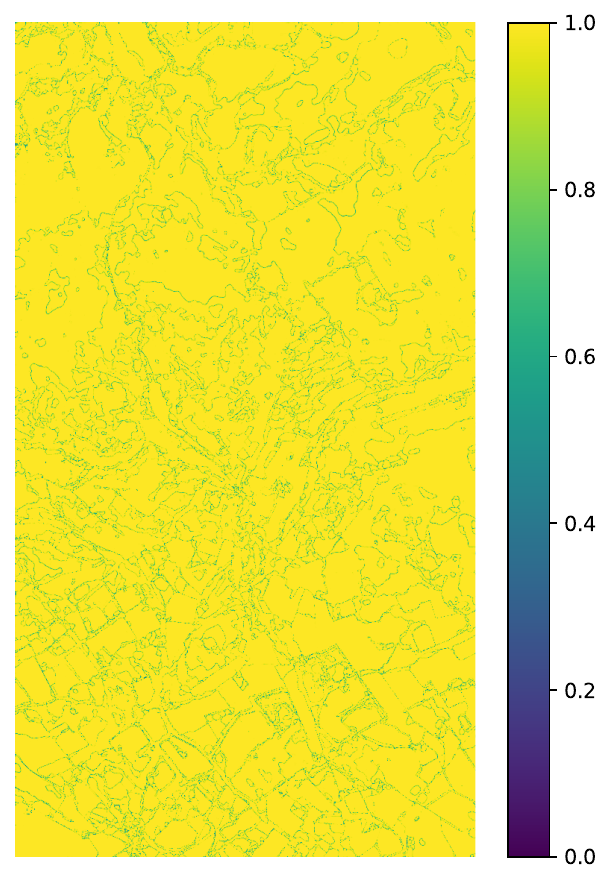}
        \caption{Probability map\label{fig:Glensaugh_result_probs}}
    \end{subfigure}
    \caption{Maps obtained from applying the POTTERS model to the Glensaugh area.}
    \label{fig:Glensaugh_result}
\end{figure}

The result from running our POTTERS model can be found in Figure~\ref{fig:Glensaugh_result}. The cluster map in Figure~\ref{fig:Glensaugh_result_class} shows the most probable class under the variational approximation to the posterior $p(z|x,\pi,\mu, \Sigma)$ from our final model.  The variational approximation further provides uncertainty quantification associated with each assigned pixel. In Figure \ref{fig:Glensaugh_result_probs}, we visualise the estimated probabilities corresponding to each pixel's most likely class label, which indicates that the cluster assignments are generally reliable, with lower posterior confidence occurring along class boundaries. The probability maps for all classes can be found in Appendix~\ref{sec:appendix_maps}.

\begin{figure}
    \centering
    \includegraphics[width=1.00\linewidth]{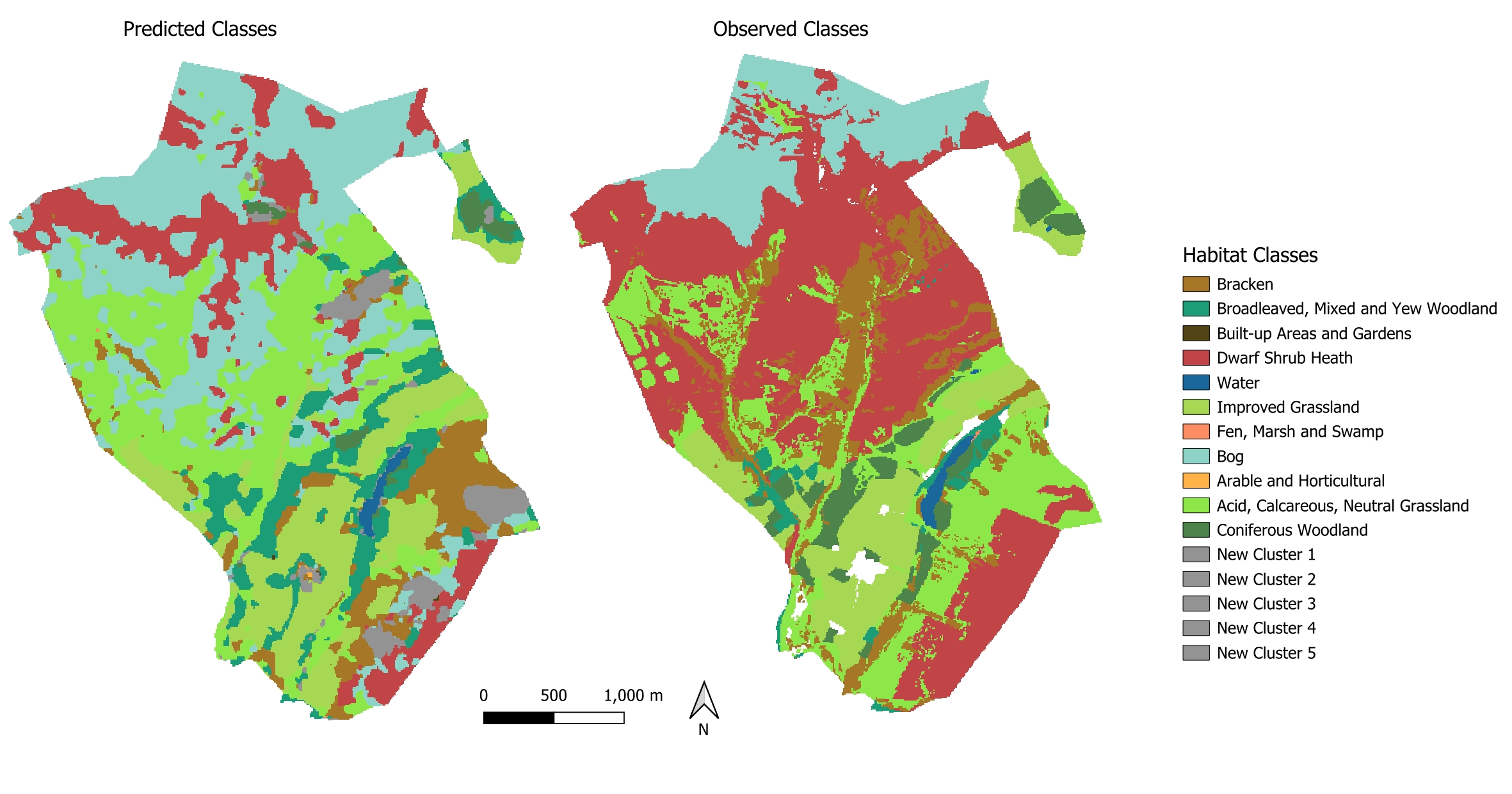}
    \caption{Comparison between the predicted classes and the ground truth labels.}
    \label{fig:validation_class_comparison}
\end{figure}

To assess the performance of our approach, we use a validation dataset obtained from an ecological survey collected by EOLAS Insight for an Innovate UK funded project. The survey used a combination of aerial photo digitisation and field-walking to collect a consistent habitat map of the study area. The habitats were mapped using the UK Habitat (UKHab) Classification System V2 (\url{https://www.ukhab.org/}) with polygons mapped to UKHab Level 4 classes. UKHab classes have a direct correspondence to the Living England classes used in the case study, meaning that each validation polygon was re-labelled with its corresponding Living England class. Figure~\ref{fig:validation_class_comparison} shows that the model can well identify superclasses such as grassland or water. Meanwhile, some subclasses, such as Improved Grassland versus Acid, Calcareous, Neutral Grassland, or Bog versus Dwarf Shrub Heath, which are difficult to distinguish from remote sensing images, contain overlaps. However, these spectrally similar subclasses are often correctly predicted by the second most probable posterior class probability, see Figure~\ref{fig:Glensaugh_result_second_class} in the Appendix, which emphasises our model's strength compared with existing clustering methods.

The overall accuracy when considering first-level classes is $47\%$, which is significantly lower than the accuracies from existing classifiers (see Figure~\ref{fig:validation_accuracy_map} for the accuracy map and Figure~\ref{fig:validation_confusion_matrix} for the confusion matrix). However, the novelty and strength of our approach is that we achieve this with no labelled training data from the target domain, which allows us to classify an area for which little or no prior knowledge is available. In comparison to other models, e.g. deep neural network models, which require some known labels for training, our approach only needs an available labelled external dataset with similar classes to explore a new area. This reduces the cost of manual labelling and improves the scalability of land cover mapping to large areas. The relatively low accuracy can be explained by a few factors. First, the satellite imagery for our labelled external data ($2019-2020$) and our target data ($2023-2024$) come from different years, which may have had different growing and blooming seasons, and varying levels of cloud cover. Second, our methodology relies on accurate labels for the external data, which are estimated to be around 87\% accurate for the Living England Habitat map we are using~\citep{kilcoyne2022living}, but which we observed is less accurate for some classes, particularly Bracken, which also affects the quality of the segmentation.  It is worth noting that Bracken is sometimes correctly identified by the second most likely class under the variational posterior, see Figure~\ref{fig:Glensaugh_result_second_class} in the Appendix. Third, heterogeneous localised patterns in either external or target data can make the problem more complicated when the characteristics of a land cover type can vary across locations. Fourth, the target test site is a challenging domain, and the accuracy would likely be higher if we considered the full Glensaugh region (pictured in Figure~\ref{fig:Glensaugh_data_spring}); however, for this, we have no ecological survey data available.

Our approach can also identify new classes: by looking at the satellite images, we notice that one new cluster appears to correspond to greenhouses, which we have not observed in our external data, see Figure~\ref{fig:prob_maps} in the Appendix for the concomitant probability map.  Moreover, this area does not belong to any known class, which explains why the algorithm places them in the new clusters.

\section{Discussion}
\label{sec:discussion}
In this work, we propose a flexible and general-purpose Bayesian image segmentation method, called POTTERS. Our model incorporates multiple novel contributions. The first novelty is the incorporation of class-dependent spatial patterns by using a generalised Potts model: class-specific hyperparameters model spatial dependencies, indicating how likely different classes may appear close to each other. Secondly, we introduce an informed prior that is estimated from an external labelled data, and a `trust' parameter that explicitly accounts for distribution shift to allow genaralisability to different images. The third novelty of our model is its ability to automatically discover new land-cover types when they are present. We allow new clusters by placing non-informative priors on model parameters for new clusters, while using constant spatial parameters for new clusters. Finally, to ensure computational scalability of our model to large remote sensing images, we propose an efficient mean-field Variational Inference algorithm for fast posterior approximation.

Similar ideas to ours have been found in other works.  The Hidden Potts mixture model by~\citep{li_bayesian_2019} allows for class-dependent spatial patterns, but is not scalable to large images due to its reliance on MCMC, and does not capture new classes nor incorporate external data. Deep neural network-based methods are scalable to large images but usually overestimate prediction accuracy ~\citep{kristiadi2020being}. Recent work on Bayesian clustering \citep{nasios_variational_2006, brodersen_variational_2013, wade_bayesian_2023} does not typically account for distribution shifts. In this context, our approach tackles all of these challenges within a single model, providing a more coherent and practical framework.

We demonstrate the benefits of our approach in numerical simulations. When dealing with the simulated data, our model can separate different clusters and find new clusters correctly; when running without any Potts parameters, some classes can be misclassified or are classified with lower certainty. Furthermore, incorporating the spatial correlations helps the result remain stable whether classes are well-separated or highly collapsed, and even when the labelled data are slightly different from the target one (i.e., when distribution shift is present).

We also applied our model to a case study of the Glensaugh area in Scotland, with the ground truth data for validation provided by our collaborators at EOLAS Insight. The results demonstrate that incorporating spatial correlation and the trust parameter yields high superclass assignment accuracy. Apart from known clusters from the external data, our approach identified new clusters and provided class probabilities for each pixel.  In addition, our approach highlights the ability to transfer domain knowledge between different locations, which reduces the need for manual labelling by experts. 

This work opens many possible directions for further research. One particular interest is having an efficient method to estimate $\Lambda^*$ under the full likelihood. Currently, we are using a maximum pseudolikelihood estimation procedure, which can be biased as it is only based on the pseudolikelihood. Recent works have started exploring Bayesian inference for intractable likelihood \citep{matsubara_generalized_2024, laplante_conjugate_2025, sonobe2026contrastive}; however, these approaches focus on binary-class settings instead of a multi-class framework. Another interesting aspect to explore is to develop variational inference methods that do not rely on the mean-field assumption. The mean-field approximation assumes that all latent variables are independent under the variational distribution, which can be a limit in the case of complex dependencies and reduce the model's generality. An alternative approach could be the fixed-form variational Bayes~\citep{tran_practical_2021} or ensemble method~\citep{evensen2003ensemble}.

\subsection*{Author contributions\label{sec:acknowledgments}}
BKN: model development, implementation of methodology, manuscript preparation. IC: data access for case study, case study validation, and manuscript checking. CB: model development, manuscript preparation. TS: model development, funding acquisition, project supervision, manuscript preparation. 

\bibliographystyle{apalike}
\bibliography{references, references_zotero}

\newpage
\appendix

\section{Appendix}
\subsection{Conditional formulation of the generalised Potts model\label{sec:conditional_formulation_potters}}

It is often useful to write the Potts model in its conditional probability formula. For the standard Potts model, this results in 
$p_\lambda(z_{n+1} \vert z_1, \ldots, z_n) \propto \exp\left( \lambda \sum_{j=1}^n W_{ij} \mathbbm{1}(z_j = z_{n+1}) \right) $. In our extended Potts model, however, we need to account for all neighbours of the $n+1$'th point, not just the ones allocated to the same cluster. We can then write
\begin{align*}
p(z_{n+1} \vert z_1, \ldots, z_n) &\propto \pi_{z_{n+1}} \exp\left( \sum_{k \in [K]} \mathbbm{1}(z_{n+1} = k) \sum_{j = 1}^n W_{ij} \lambda_{k,z_{j}} \right) =\\
&=\pi_{z_{n+1}} \exp\left( \sum_{k \in [K]} \mathbbm{1}(z_{n+1} = k)  \sum_{h \in [K]} \lambda_{k,h} \sum_{j = 1}^n W_{ij} \mathbbm{1}(z_j = h) \right),
\end{align*}
where $\sum_{j = 1}^n W_{ij} \mathbbm{1}(z_j = h)$ 
counts how many neighbors of the $n+1$th element belong to each of the clusters $h$.
Note: if $\lambda_{k,h} = C$ for all $k,h$, then we would have a uniform probability of choosing any of the clusters, the same as if $\lambda_{k,h} = 0$.  The same can also be written as
\begin{equation}
  p(z_{n+1} \vert z_1, \ldots, z_n) \propto
    \begin{cases}
      \pi_1 \cdot \exp \left( \sum_{h} \lambda_{1,h}  \sum_{j = 1}^n W_{(n+1)j} \mathbbm{1}(z_j = h) \right) & \text{if $z_{n+1} = 1$}\\
      \ldots\\
      \pi_K \cdot \exp \left( \sum_{h} \lambda_{K,h}  \sum_{j = 1}^n W_{(n+1)j} \mathbbm{1}(z_j = h) \right) & \text{if $z_{n+1} = K$}
    \end{cases}       .
\end{equation}

\subsection{Matrix formulation of the generalised Potts model\label{sec:matrix_formulation_potters}}

Let $Z$ denote the $n \times K$ matrix that encodes the cluster allocations, with $Z_{ik} = 1$ if $z_i = k$, i.e. if element $i$ belongs to cluster $k$. Let $W$ be the $n \times n$ adjacency matrix, with $W_{ij}=1$ if elements $i$ and $j$ are adjacent, and 0 otherwise. Set also $W_{ii} = 0$ by default. Finally, let $\Lambda$ be a $K \times K$ matrix containing the collection of $\lambda$-parameters in the model. 

We can rewrite $M_\Lambda(z)$ as 
$$
M_\Lambda(z) \propto \exp\left(- \frac12 \textrm{tr}(\Lambda^T (Z^T WZ) )\right).
$$

First, note that 
\begin{align*}
(Z^T WZ)_{k, h} &= \sum_{i} (Z^T)_{k i} (WZ)_{i h} = \sum_{i} Z_{ik} (WZ)_{ih} = \\
& = \sum_{i} Z_{ik} \sum_{j} W_{ij} Z_{jh} = \sum_{i,j} Z_{ik} W_{ij} Z_{jh} = \\
& = \sum_{i,j} W_{ij} \mathbbm{1} \{z_i=k,z_j=h\} = 2 \sum_{i<j \in [n]} W_{ij} \mathbbm{1} \{z_i=k,z_j=h\}.
\end{align*}

Remember also that $\textrm{tr}(A^TB)$ is equal to the sum of the elements of the element-wise (Hadamard) product of two matrices, $A$ and $B$. Thus, 
\begin{align*}
\textrm{tr}(\Lambda^T (Z^T WZ)) &= \sum_{k, h} \lambda_{k,h} (Z^T WZ)_{k,h} = \\
&= 2\sum_{k, h} \sum_{i<j \in [n]} W_{ij} \mathbbm{1} \{z_i=k,z_j=h\}.
\end{align*}

\subsection{Proof for Theorem \ref{thm: thm1}}\label{sec:thm_proof}
Let $\widehat{\Lambda}$ be the solution of the optimisation problem (\ref{eq:optimisation_problem}), i.e. 
\[
\widehat{\Lambda} \in \argmax_{\Lambda} \{\tilde{L}(z; \Lambda)\},
\]
or, equivalently, 
\[
\tilde{L}(z;\widehat{\Lambda}) \geq  \tilde{L}(z; \Lambda), \quad \text{for any $\Lambda$},
\]
where we note that $\tilde{L}(z;\widehat{\Lambda})\neq 0$.
Suppose now that $\widehat{\Lambda}$ is such that $\widehat{\lambda}_{kh}\in\IR$, then
\begin{align*}
\tilde{L}(z; 2\widehat{\Lambda}) 
&\propto \displaystyle \sum_{i=1}^{N} \log \Biggl\{ \dfrac{ \exp \Big( \displaystyle \sum_{j \in N(i)}  \displaystyle \sum_{h=1}^{\tilde{K}} 2\widehat{\lambda}_{kh} \mathbbm{1}\{z_i \neq z_j, z_i = k, z_j = h \} \Bigr) }{ \displaystyle \sum_{k'=1}^{\tilde{K}} \exp \Big( \sum_{j \in N(i)}  \displaystyle \sum_{h=1}^{\tilde{K}} 2\widehat{\lambda}_{k'h} \mathbbm{1}\{z_i \neq z_j, z_i = k', z_j = h \} \Bigr) } \Biggr\} 
    \\
    &= \displaystyle \sum_{i=1}^{N} \log\Biggl\{ \dfrac{ \Big[ \exp \Big( \displaystyle \sum_{j \in N(i)}  \displaystyle \sum_{h=1}^{\tilde{K}}\widehat{\lambda}_{kh} \mathbbm{1}\{z_i \neq z_j, z_i = k, z_j = h \} \Bigr) \Big]^2 }{ \displaystyle \sum_{k'=1}^{\tilde{K}} \Big[\exp \Big( \sum_{j \in N(i)}  \displaystyle \sum_{h=1}^{\tilde{K}} \widehat{\lambda}_{k'h} \mathbbm{1}\{z_i \neq z_j, z_i = k', z_j = h \} \Bigr) \Big]^2 } \Bigg\} 
    \\
    &= \displaystyle \sum_{i=1}^{N} 2 \log \Biggl\{ \dfrac{ \exp \Big( \displaystyle \sum_{j \in N(i)}  \displaystyle \sum_{h=1}^{\tilde{K}} \widehat{\lambda}_{kh} \mathbbm{1}\{z_i \neq z_j, z_i = k, z_j = h \} \Bigr) }{ \displaystyle \sum_{k'=1}^{\tilde{K}} \exp \Big( \sum_{j \in N(i)}  \displaystyle \sum_{h=1}^{\tilde{K}} \widehat{\lambda}_{k'h} \mathbbm{1}\{z_i \neq z_j, z_i = k', z_j = h \} \Bigr) } \Biggr\} 
    \\
    &=2 \tilde{L}(z;\widehat{\Lambda}) > \tilde{L}(z;\widehat{\Lambda}), 
\end{align*}
so $\widehat{\Lambda}$ cannot be a maximiser, the proof concludes by contradiction.

\subsection{Variational Updates} \label{sec:appendix variational distribution}
\subsubsection{Update \texorpdfstring{$c$}{c}}
Recall that the optimal solution for the ELBO optimisation problem is 
\begin{equation}
    \ln q^*(z_i) = \mathbb{E}_{-z_i}[\ln p(x,z, \mu, \Sigma)],
\end{equation} 
which means we find the optimal variational factor by taking the expectation of the likelihood with respect to every parameter except the parameter of the variational factor we are considering.

Consider the variational distribution $q^*(z)$
\begin{align*}
    \ln q^*(&z_i) = \mathbb{E}_{-z_i}\bigr[\ln p(x|z,\mu,\Sigma) + \ln p(z|\pi, \Lambda^*) \bigl]
    \\
    &= \mathbb{E}_{-z_i}\Bigr[\sum_k \mathbbm{1}\{z_i = k\} \ln \mathcal{N}(x_i|\mu_k, \Sigma_k) +  \sum_k \mathbbm{1}\{z_i = k\}\ln \pi_k 
    \\
    & \hspace{12mm} + \sum_{k=1}^K\sum_{j \in N(i)} \sum_{h=1}^K \lambda^*_{kh}\mathbbm{1}\{z_i = k, z_j = h, z_i \neq z_j \} \Bigl] 
    \\
    &= \mathbb{E}_{-z_i} \Bigr[ \sum_k \mathbbm{1}\{z_i = k\} \Big(-\dfrac{1}{2} \ln |\Sigma_k| - \dfrac{1}{2}(x_i-\mu_k)^T\Sigma_k^{-1}(x_i- \mu_k) - \frac{d}{2}\ln 2\pi \Big) 
    \\
    & \hspace{12mm} + \sum_k \mathbbm{1}\{z_i = k\} \ln \pi_k + \sum_{k=1}^K\sum_{j \in N(i)} \sum_{h=1}^K \lambda^*_{kh} \mathbbm{1}\{z_i = k, z_j = h, z_i \neq z_j\} \Bigl]
    \\
    & = \sum_k \mathbbm{1}\{z_i = k\}\Bigg\{-\frac{1}{2}  \mathbb{E}(\ln|\Sigma_k|) -\frac{1}{2} \mathbb{E}[(x_i-\mu_k)^T\Sigma_k^{-1}(x_i - \mu_k)]) - \frac{d}{2}\ln 2\pi  + \ln \pi_k \Bigg\} 
    \\
    & \hspace{12mm} + \mathbb{E}\Big\{\sum_{k=1}^K\sum_{j \in N(i)} \sum_{h=1, h\neq k}^K \lambda^*_{kh} \mathbbm{1}\{z_i = k\} \mathbbm{1}\{z_j = h\} \Big\}
    \\
    & = \sum_k \mathbbm{1}\{z_i = k\} \Biggl \{ -\frac{1}{2} \mathbb{E}(\ln|\Sigma_k|) -\frac{1}{2} \mathbb{E}[(x_i\!-\!\mu_k)^T\Sigma_k^{-1}(x_i\!-\!\mu_k)] - \frac{d}{2}\ln 2\pi + \ln \pi_k
    \\
    & \hspace{32mm} + \sum_{j \in N(i)} \sum_{h=1}^K \lambda^*_{kh} c_{jh}  \Biggr\},
\end{align*}

where 
\begin{align}\begin{split}
    \mathbb{E}[\ln |\Sigma_k|] &=  -d\ln 2 + \ln |C_k| - \sum_{i=1}^d \psi \Bigr( \dfrac{v_k + 1 - i}{2} \Bigl)
    \\
    \mathbb{E}[(x_i-\mu_k)^T\Sigma_k^{-1}(x_i- \mu_k)]& = v_k(x_i - m_k)^T C_k^{-1} (x_i - m_k) + db_k^{-1}.
    \label{eq: common expectation terms}
\end{split}\end{align}
Now, we set 
\begin{equation}
    \ln c'_{ik} = -\frac{1}{2} \mathbb{E}(\ln|\Sigma_k|) -\frac{1}{2} \mathbb{E}[(x_i -\mu_k)^T\Sigma_k^{-1}(x_i - \mu_k)] - \frac{d}{2}\ln 2\pi + \ln \pi_k + \sum_{j \in N(i)} \sum_{h=1}^K \lambda^*_{kh} c_{jh},
\end{equation}
and,
\begin{equation}
    c_{ik} = \dfrac{c'_{ik}}{\displaystyle \sum_{l=1}^K c'_{il}}. 
\end{equation}
Then, the optimal $q^*(z)$ follows the Categorial distribution
\[ 
q^*(z) \propto \prod_i \prod_k c_{ik}^{\mathbbm{1}\{z_i = k\}}. 
\]

\subsubsection{Update \texorpdfstring{$m, b, v, C$}{m, b, v, C}}

Finally, we prove that the variational distribution for $q^*(\mu_k, \Sigma_k)$ is the Normal Inverse Wishart distribution 
\[
q^*(\mu_k, \Sigma_k) = \mathcal{NIW}(\mu_k, \Sigma_k|m_k, b_k, v_k, C_k),
\]
where the updates for variational parameters are defined as 

\begin{align}\begin{split}
    b_k &= 
    \begin{cases}
        \tilde{\beta}_k + N_k &\text{for} \quad 1\leq k \leq \tilde{K} 
        \\
        \kappa + N_k &\text{for} \quad \tilde{K} +1 \leq k \leq K
    \end{cases}
    \\
    m_k &= 
    \begin{cases}
        \dfrac{\tilde{\beta}_k \tilde{\eta}_k + N_k \bar{x}_k}{N_k+ \tilde{\beta}_k} &\text{for} \quad 1\leq k \leq \tilde{K}  
        \\
        \\
        \dfrac{\kappa \eta + N_k \bar{x}_k}{N_k + \kappa} &\text{for} \quad \tilde{K} +1 \leq k \leq K 
    \end{cases}
    \\
    v_k &= 
    \begin{cases}
        \tilde{\nu}^{\text{adj}}_k + N_k &\text{for} \quad 1\leq k \leq \tilde{K} 
        \\
        \nu + N_k &\text{for} \quad \tilde{K} +1 \leq k \leq K
    \end{cases}
    \\
    C_k &= 
    \begin{cases}
        \tilde{\Psi}^{\text{adj}}_k + N_k S_k + \dfrac{N_k \tilde{\beta}_k}{N_k+ \tilde{\beta}_k}  (\bar{x}_k - \tilde{\eta}^{\text{adj}}_k)(\bar{x}_k - \tilde{\eta}^{\text{adj}}_k)^T &\text{for} \quad 1\leq k \leq \tilde{K} 
        \\
        \\
        \Psi + N_k S_k + \dfrac{N_k \kappa}{N_k + \kappa}  (\bar{x}_k - \eta)(\bar{x}_k - \eta)^T &\text{for} \quad \tilde{K} +1 \leq k \leq K.
    \end{cases}
    \label{eq: update mu, Sigma}
\end{split}\end{align}

Consider the $q^*(\mu_k, \Sigma_k)$ in 2 cases: for known clusters when $k = 1, \ldots \tilde{K}$ and for unknown clusters when $k = \tilde{K} + 1, \ldots, K$. In the first case, we can write the optimal variational distribution as 
\begin{align*}
    \ln q^*(\mu_k, \Sigma_k) &= \mathbb{E}_{-\mu_k, \Sigma_k} \bigr[ \ln p(x|z, \mu, \Sigma) + \ln p(\mu, \Sigma) \big]
    \\
    & \propto \mathbb{E}_{-\mu_k, \Sigma_k} \Big[\sum_i \mathbbm{1}\{z_i = k\} \Big(-\frac{1}{2} \ln |\Sigma_k| - \frac{1}{2} (x_i - \mu_k)^T \Sigma_k^{-1} (x_i - \mu_k) \Big)
    \\
    &\hspace{18mm} - \frac{1}{2} \ln |\Sigma_k| - \frac{1}{2} (\mu_k - \tilde{\eta}_k)^T (\tilde{\beta}_k \Sigma_k^{-1}) (\mu_k - \tilde{\eta}_k)
    \\
    &\hspace{18mm} - \frac{\tilde{\nu}^{\text{adj}}_k}{2} \ln |\tilde{\Psi}^{\text{adj}}_k| - \frac{\tilde{\nu}^{\text{adj}}_k + d + 1}{2} \ln |\Sigma_k| - \frac{1}{2} \operatorname{tr}(\tilde{\Psi}^{\text{adj}}_k \Sigma_k^{-1})
    \Big] 
    \\
    &=\mathbb{E}_{-\mu_k, \Sigma_k} \Big\{ \ln |\Sigma_k| \cdot \left(-\dfrac{\sum_i \mathbbm{1}\{z_i = k\} + \tilde{\nu}^{\text{adj}}_k + d + 2}{2} \right) \\
    & \hspace{20mm} - \dfrac{1}{2} \Big[\sum_i \mathbbm{1}\{z_i = k\} (x_i - \mu_k)^T \Sigma_k^{-1} (x_i - \mu_k) 
    \\
    & \hspace{28mm} + (\mu_k - \tilde{\eta}_k)^T \tilde{\beta}_k \Sigma_k^{-1} (\mu_k - \tilde{\eta}_k) + \operatorname{tr}(\tilde{\Psi}^{\text{adj}}_k \Sigma_k^{-1}) \Big] \Big\}.
\end{align*}
Taking the expectation with respect to the variational distribution of every parameter except $\mu_k, \Sigma_k$, we get 
\begin{align*}
    \ln q^*(\mu_k, \Sigma_k) &= \ln |\Sigma_k| \Big(-\dfrac{N_k + \tilde{\nu}^{\text{adj}}_k + d + 2}{2} \Big) 
    \\ 
    &- \dfrac{1}{2} \Big( N_k \big[(\bar{x}_k\!-\! \mu_k)^T\Sigma_k^{-1}(\bar{x}_k\!-\!\mu_k) \big] + (\mu_k\!-\! \tilde{\eta}_k)^T \tilde{\beta}_k\Sigma_k^{-1}(\mu_k\!-\! \tilde{\eta}_k) + \operatorname{tr}(\tilde{\Psi}^{\text{adj}}_k \Sigma_k^{-1}) \Big) 
    \\
    &=: \mathrm{Term}_1 - \dfrac{1}{2}\mathrm{Term}_2.
\end{align*}
The first term gives us the update for $\tilde{\nu}_k$. Consider the second term, we have 
\begin{align}
    \mathrm{Term}_2 &= \sum_i x_i^T\Sigma_k^{-1}x_i - N_k \bar{x}_k\Sigma_k^{-1}\mu_k - N_k \mu_k^T\Sigma_k^{-1} \bar{x}_k + N_k \mu_k^T \Sigma_k^{-1}\mu_k \notag
    \\
    &\hspace{6mm} + \mu_k^T \tilde{\beta}_k\Sigma_k^{-1}\mu_k - \mu_k^T \tilde{\beta}_k\Sigma_k^{-1}\tilde{\eta}_k - \tilde{\eta}_k^T \tilde{\beta}_k\Sigma_k^{-1}\mu_k + \tilde{\eta}_k^T \tilde{\beta}_k \Sigma_k^{-1}\tilde{\eta}_k + \operatorname{tr}(\tilde{\Psi}^{\text{adj}}_k\Sigma_k^{-1}) \notag
    \\
    & = (N_k + \tilde{\beta}_k) \mu_k^T \Sigma_k^{-1} \mu_k - \mu_k^T \Sigma_k^{-1}(N_k \bar{x}_k + \tilde{\beta}_k\tilde{\eta}_k) - (N_k \bar{x}_k + \tilde{\beta}_k\tilde{\eta}_k)\Sigma_k^{-1}\mu_k \notag
    \\
    &\hspace{6mm} + \tilde{\beta}_k\tilde{\eta}_k^T \Sigma_k^{-1}\tilde{\eta}_k + \sum_i x_i^T \Sigma_k^{-1}x_i + \operatorname{tr}(\tilde{\Psi}^{\text{adj}}_k \Sigma_k^{-1}) \notag
    \\
    & = (N_k + \tilde{\beta}_k) \mu_k^T \Sigma_k^{-1} \mu_k - \mu_k^T \Sigma_k^{-1}(N_k \bar{x}_k + \tilde{\beta}_k\tilde{\eta}_k) - (N_k \bar{x}_k + \tilde{\beta}_k \tilde{\eta}_k)\Sigma_k^{-1}\mu_k \notag
    \\ 
    &\hspace{6mm} + \dfrac{1}{N_k + \tilde{\beta}_k} (\tilde{\beta}_k\tilde{\eta}_k + N_k \bar{x}_k)^T \Sigma_k^{-1}(\tilde{\beta}_k\tilde{\eta}_k + N_k \bar{x}_k) \notag
    \\
    &\hspace{6mm} - \dfrac{1}{N_k + \tilde{\beta}_k} (\tilde{\beta}_k\tilde{\eta}_k + N_k \bar{x}_k)^T \Sigma_k^{-1}(\tilde{\beta}_k\tilde{\eta}_k + N_k \bar{x}_k) \notag
    \\
    &\hspace{6mm} + \tilde{\beta}_k\tilde{\eta}_k^T \Sigma_k^{-1}\tilde{\eta}_k + \sum_i x_i^T \Sigma_k^{-1}x_i + \operatorname{tr}(\tilde{\Psi}^{\text{adj}}_k \Sigma_k^{-1}).
    \label{eq: term2}
\end{align}
In the equation (\ref{eq: term2}), the first 2 lines give us
\[
(\tilde{\beta}_k + N_k) \Bigg( \mu_k - \dfrac{\tilde{\beta}_k\tilde{\eta}_k + N_k \bar{x}_k}{N_k + \tilde{\beta}_k}\Bigg)^T \Sigma_k^{-1} \Bigg( \mu_k - \dfrac{\tilde{\beta}_k\tilde{\eta}_k + N_k \bar{x}_k}{N_k + \tilde{\beta}_k}\Bigg).
\]
Consider the last 2 lines, adding and subtracting the term $N_k \bar{x}_k \Sigma_k^{-1}\bar{x}_k$, we get 
\begin{align}\begin{split}
    & \sum_i (x_i^T \Sigma_k^{-1} x_i - x_i^T \Sigma_k^{-1}\bar{x}_k - \bar{x}^T\Sigma_k^{-1}x_i + \bar{x}_k^T \Sigma_k^{-1} \bar{x}_k) 
    \\
    & \hspace{2mm}  +N_k \bar{x}_k \Sigma_k^{-1}\bar{x}_k + \tilde{\beta}_k\tilde{\eta}_k^T \Sigma_k^{-1}\tilde{\eta}_k  
    \\
    & \hspace{2mm} - \dfrac{1}{N_k + \tilde{\beta}_k} (\tilde{\beta}_k\tilde{\eta}_k + N_k \bar{x}_k)^T \Sigma_k^{-1}(\tilde{\beta}_k\tilde{\eta}_k + N_k \bar{x}_k) + \operatorname{tr}(\tilde{\Psi}^{\text{adj}}_k\Sigma_k^{-1}).
    \label{eq: last 2 lines}
\end{split}\end{align}
The first line of (\ref{eq: last 2 lines}) gives us 
\[
\sum_i (x_i - \bar{x}_k)^T\Sigma_k^{-1}(x_i - \bar{x}_k).
\]
The next two lines of (\ref{eq: last 2 lines}) can be expanded to 
\begin{align*}
    & N_k \bar{x}_k \Sigma_k^{-1}\bar{x}_k \! + \! \tilde{\beta}_k\tilde{\eta}_k^T \Sigma_k^{-1}\tilde{\eta}_k \!+\! \operatorname{tr}(\tilde{\Psi}^{\text{adj}}_k\Sigma_k^{-1})
    \\
    & \hspace{8mm}- \dfrac{1}{N_k + \tilde{\beta}_k} (\tilde{\beta}_k^2 \tilde{\eta}_k^T \Sigma_k^{-1}\tilde{\eta}_k \!+\! N_k \tilde{\beta}_k\bar{x}_k^T \Sigma_k^{-1} \tilde{\eta}_k \!+\! N_k \tilde{\beta}_k \tilde{\eta}_k^T \Sigma_k^{-1} \bar{x}_k) 
    \\
    = & \dfrac{N_k \tilde{\beta}_k}{N_k + \tilde{\beta}_k} (\bar{x}_k^T \Sigma_k^{-1} \bar{x} -  \bar{x}_k^T \Sigma_k^{-1} \tilde{\eta}_k - \tilde{\eta}_k^T \Sigma_k^{-1} \bar{x}_k + \tilde{\eta}_k^T \Sigma_k^{-1}\tilde{\eta}_k)  + \operatorname{tr}(\tilde{\Psi}^{\text{adj}}_k \Sigma_k^{-1})
    \\
    = & \dfrac{N_k \tilde{\beta}_k}{N_k + \tilde{\beta}_k}  (\bar{x}_k - \tilde{\eta}_k)^T \Sigma_k^{-1}(\bar{x}_k - \tilde{\eta}_k)  + \operatorname{tr}(\tilde{\Psi}^{\text{adj}}_k\Sigma_k^{-1}).
\end{align*}
Therefore, the \ref{eq: last 2 lines} can be fully written as
\begin{align*}
    &\operatorname{tr} \Bigg[ (x_i - \bar{x}_k)^T\Sigma_k^{-1}(x_i - \bar{x}_k) + \dfrac{N_k \tilde{\beta}_k}{N_k + \tilde{\beta}_k}  (\bar{x}_k - \tilde{\eta}_k)^T \Sigma_k^{-1}(\bar{x}_k - \tilde{\eta}_k) + \tilde{\Psi}^{\text{adj}}_k\Sigma_k^{-1}   \Bigg] 
    \\
    & =  \operatorname{tr} \Bigg[(x_i - \bar{x}_k)(x_i - \bar{x}_k)^T\Sigma_k^{-1} + \dfrac{N_k \tilde{\beta}_k}{N_k + \tilde{\beta}_k}  (\bar{x}_k - \tilde{\eta}_k)(\bar{x}_k - \tilde{\eta}_k)^T \Sigma_k^{-1} + \tilde{\Psi}^{\text{adj}}_k\Sigma_k^{-1}  \Bigg] 
    \\
    & = \operatorname{tr} \Bigg[ \Bigg( (x_i - \bar{x}_k)(x_i - \bar{x}_k)^T + \dfrac{N_k \tilde{\beta}_k}{N_k + \tilde{\beta}_k}  (\bar{x}_k - \tilde{\eta}_k)(\bar{x}_k - \tilde{\eta}_k)^T + \tilde{\Psi}^{\text{adj}}_k \Bigg)\Sigma_k^{-1} \Bigg].
\end{align*}

Then, 
\begin{align*}
    \ln q^*(\mu_k, \Sigma_k) &= \ln |\Sigma_k| \Big(-\dfrac{N_k + \tilde{\nu}^{\text{adj}}_k + d + 2}{2} \Big) 
    \\
    &\hspace{6mm} -\dfrac{1}{2} (\tilde{\beta}_k + N_k) \Bigg( \mu_k - \dfrac{\tilde{\beta}_k\tilde{\eta}_k + N_k \bar{x}_k}{N_k + \tilde{\beta}_k}\Bigg)^T \Sigma_k^{-1} \Bigg( \mu_k - \dfrac{\tilde{\beta}_k\tilde{\eta}_k + N_k \bar{x}_k}{N_k + \tilde{\beta}_k}\Bigg)
    \\
    &\hspace{6mm} - \dfrac{1}{2} \operatorname{tr} \Bigg[ \Bigg(\tilde{\Psi}^{\text{adj}}_k + (x_i - \bar{x}_k)(x_i - \bar{x}_k)^T + \dfrac{N_k \tilde{\beta}_k}{N_k + \tilde{\beta}_k}  (\bar{x}_k - \tilde{\eta}_k)(\bar{x}_k - \tilde{\eta}_k)^T \Bigg)\Sigma_k^{-1} \Bigg]
    \\
    & = \ln |\Sigma_k| \Big(-\dfrac{N_k + \tilde{\nu}^{\text{adj}}_k + d + 2}{2} \Big) 
    \\
    &\hspace{6mm} - \dfrac{1}{2}(\tilde{\beta}_k+ N_k) \Bigg( \mu_k - \dfrac{\tilde{\beta}_k\tilde{\eta}_k + N_k \bar{x}_k}{N_k + \tilde{\beta}_k}\Bigg)^T \Sigma_k^{-1} \Bigg( \mu_k - \dfrac{\tilde{\beta}_k\tilde{\eta}_k + N_k \bar{x}_k}{N_k + \tilde{\beta}_k^{-1}}\Bigg)
    \\
    &\hspace{6mm}-\dfrac{1}{2} \operatorname{tr} \Bigg[ \Bigg( \tilde{\Psi}^{\text{adj}}_k + N_k S_k + \dfrac{N_k \tilde{\beta}_k}{N_k + \tilde{\beta}_k}  (\bar{x}_k - \tilde{\eta}_k)(\bar{x}_k - \tilde{\eta}_k)^T \Bigg)\Sigma_k^{-1} \Bigg],
\end{align*}
which gives us the variational distribution $q^*(\mu_k, \Sigma_k) = \mathcal{NIW}(m_k, b_k, v_k, C_k)$. These variational parameters are defined as
\begin{align*}
    b_k &= \tilde{\beta}_k + N_k   
    \\
    m_k &= \dfrac{\tilde{\beta}_k \tilde{\eta}_k + N_k \bar{x}_k}{N_k+ \tilde{\beta}_k}
    \\
    v_k &= \tilde{\nu}^{\text{adj}}_k + N_k 
    \\
    C_k &= \tilde{\Psi}^{\text{adj}}_k + N_k S_k + \dfrac{N_k \tilde{\beta}_k}{N_k + \tilde{\beta}_k}  (\bar{x}_k - \tilde{\eta}_k)(\bar{x}_k - \tilde{\eta}_k)^T.
\end{align*}

For the second case, when considering the new cluster, the calculation is the same as before, except that the prior parameters are different. Then, the updates for variational parameters for new clusters ($k = \tilde{K}, \ldots, K$) will be 
\begin{align*}
    b_k &= \kappa + N_k 
    \\
    m_k &= \dfrac{\kappa \eta + N_k \bar{x}_k}{N_k + \kappa} 
    \\
    v_k &= \nu + N_k 
    \\
    C_k &=  \Psi + N_k S_k + \dfrac{N_k \kappa}{N_k + \kappa}  (\bar{x}_k - \eta)(\bar{x}_k - \eta)^T .
\end{align*}

\subsection{ELBO Calculation}\label{sec:ELBO}
By \cite{blei_variational_2017}, the optimal solution for the ELBO optimisation problem is 
\begin{equation}
    \ln q^*(z_i) = \mathbb{E}_{-z_i}[\ln p(x,z, \mu, \Sigma)].
    \label{eq: optimal variational solution}
\end{equation} 

It can be written in our problem as
\begin{align}\begin{split}
    \label{eq:ELBO}
     \ELBO\bigl(q(c,b,m,v,C)\bigr) &= \mathbb{E}[\log p(X|z, \mu, \Sigma)] 
     \\
     & \hspace{3mm} + \mathbb{E}[\log p(z)] + \mathbb{E}[\log p(\mu, \Sigma)]  \\
    &\hspace{3mm} 
     - \mathbb{E}[\log q(z)] - \mathbb{E}[\log q(\mu, \Sigma)],
\end{split}
\end{align}
where the expectations are taken with respect to the variational distribution $q$, and the individual terms can be calculated as

\begin{align}\begin{split}
\mathbb{E}[\log p(X|z, \mu, \Sigma)] &= \dfrac{1}{2} \displaystyle \sum_{k=1}^{K} N_k \Big\{-d \ln 2\pi - \mathbb{E} (\ln |\Sigma_k|) - v_k (\bar{x}_k - m_k)^T C_k^{-1} (\bar{x}_k - m_k) 
\\
& \hspace{70mm} - d b_k^{-1} - v_k Tr(S_k C_k^{-1}) \Big\}
\\
\mathbb{E}[\log p(z|\pi)]& = \displaystyle \sum_{i=1}^N \sum_{k=1}^{K} c_{ik} \ln \pi_k + \sum_{i=1}^N \sum_{j \in N(i)} \sum_{k=1}^{K} \sum_{h=1}^{K} \lambda^*_{kh} c_{ik} c_{jh}
\\
\mathbb{E}[\log p(\mu, \Sigma)] &=  \sum_{k=1}^{\tilde{K} } - \dfrac{\tilde{\nu}^{\text{adj}}_k + d + 2}{2} \mathbb{E}(\ln |\Sigma_k|) - \dfrac{1}{2} v_k \operatorname{tr}(\tilde{\Psi}^{\text{adj}}_k C_k^{-1}) + \dfrac{d}{2}\ln \Big(\frac{\tilde{\beta}_k}{2\pi}\Big) 
\\
& \hspace{8mm} - \dfrac{\tilde{\beta}_k}{2} (m_k - \tilde{\eta}_k)^T v_k C_k^{-1} (m_k - \tilde{\eta}_k) - \dfrac{d \tilde{\beta}_k}{2 b_k} +  \ln B(\tilde{\Psi}_k, \tilde{\nu}_k)
\\
&+ \sum_{k=\tilde{K}+1}^{K}-\dfrac{\nu+ d + 2}{2} \mathbb{E}(\ln |\Sigma_k|) - \dfrac{1}{2} v_k \operatorname{tr}(\Psi C_k^{-1}) + \dfrac{d}{2}\ln \Big(\frac{\kappa}{2\pi}\Big) 
\\
& \hspace{8mm} - \dfrac{\kappa}{2} (m_k - \eta)^T v_k C_k^{-1} (m_k - \eta) - \dfrac{d \kappa}{2 b_k} +  \ln B(\Psi, \nu)
\\
\mathbb{E}[\log q(z)] &= \sum_{i=1}^N \sum_{k=1}^{K} c_{ik} \ln c_{ik}.
\\
\mathbb{E}[\log q(\mu, \Sigma)] &= \sum_{k=1}^K \Bigg\{ H\Big[q(\Sigma_k)\Big] - \dfrac{1}{2}\mathbb{E}(\ln |\Sigma_k|) + \dfrac{d}{2}\ln \Big(\frac{b_k}{2\pi} \Big) - \dfrac{d}{2} \Bigg\},
\label{eq: detailed elbo}
\end{split}\end{align} 
where $B(\cdot)$ is defined as
\begin{equation}
B(\Psi_k, \nu_k) = |\tilde{\Psi}^{\text{adj}}_k|^{\frac{\tilde{\nu}^{\text{adj}}_k}{2}} \cdot \Big(2^{\frac{\tilde{\nu}^{\text{adj}}_k d}{2}} \pi^{\frac{d(d-1)}{4}}  \prod_k \Gamma \Big(\dfrac{\tilde{\nu}^{\text{adj}}_k + 1 - i}{2} \Big)\Big) ^{-1}.
\label{eq: function B}
\end{equation}
$H(\cdot)$ is the entropy of the Inverse-Wishart distribution calculated as 
\begin{align}\begin{split}
    H\Big[q(\Sigma_k)\Big] &= \mathbb{E}\big(\log q(\Sigma_k) \big) 
    \\
    & = \mathbb{E} \Big( \log B(\Psi_k, \nu_k) - \dfrac{\nu_k + d + 1}{2}\log |\Sigma_k| - \dfrac{1}{2}\operatorname{tr}(\Psi_k \Sigma_k^{-1}) \Big)
    \\
    &= \ln B(C_k, v_k) - \dfrac{v_k + d + 1}{2}\mathbb{E}(\ln |\Sigma_k|) - \dfrac{v_k d}{2}.
\label{eq: function H} 
\end{split}\end{align}
Note that $N_k$, $\bar{x}_k$, and $S_k$ are calculated as in (\ref{eq: common terms}).
To see these equalities in detail, we in turn look at each of these seven expectations.

Consider first $\mathbb{E}[\ln p(X|z, \mu, \Sigma)]$, where we have
\begin{align*}
    \mathbb{E}[\ln p(X|z, \mu, \Sigma)] &= \mathbb{E}\Bigg[ \sum_{i=1}^n  \mathbbm{1}\{z_i = k\} \Big( \!-\!\dfrac{d}{2} \ln 2\pi \!-\! \dfrac{1}{2} \ln |\Sigma_k| \!-\! \dfrac{1}{2}(x_i \!-\! \mu_k)^T\Sigma_k^{-1}(x_i \!-\! \mu_k) \Big) \Bigg]
    \\
    &= \sum_k N_k \Big\{ -\dfrac{d}{2} \ln 2\pi - \dfrac{1}{2} \mathbb{E}(\ln |\Sigma_k|) \Big\}
    \\
    & \hspace{5mm} - \dfrac{1}{2} \mathbb{E} \left[ \mathbbm{1}(z_i = k)\Big(\sum_i x_i^T \Sigma_k^{-1}x_i - \sum_i 2 x_i^T \Sigma_k^{-1}\mu_k + \sum_i \mu_k^T \Sigma_k^{-1}\mu_k \Big)  \right]
    \\
    &= \sum_k N_k \Big\{ -\dfrac{d}{2} \ln 2\pi - \dfrac{1}{2} \mathbb{E}(\ln |\Sigma_k|) \Big\}  - \dfrac{1}{2} A_k.
\end{align*}
Consider A separately, we have  (in the second line we take expectation with respect to $z_i$
\begin{align*}
    A_k &= \mathbb{E} \Big( \sum_i \mathbbm{1}(z_i = k) (\bar{x}_k + (x_i - \bar{x}_k))^T \Sigma_k^{-1}(\bar{x}_k + (x_i - \bar{x}_k)) 
    \\
    &\hspace{32mm} - 2 \sum_i \mathbbm{1}(z_i = k) x_i \Sigma_k^{-1} \mu_k + 
    \sum_i \mathbbm{1}(z_i = k) \mu_k^T \Sigma_k^{-1} \mu_k \Big) 
    \\
    &= \mathbb{E} \Bigg\{\sum_i \mathbbm{1}(z_i = k) \Big[ \bar{x}_k^T \Sigma_k^{-1} \bar{x}_k \!+\! 2 \bar{x}_k^T \Sigma_k^{-1}(x_i \!-\! \bar{x}_k) + (x_i \!-\! \bar{x}_k)^T \Sigma_k^{-1} (x_i \!-\! \bar{x}_k) 
    \\
    & \hspace{32mm} - 2x_i^T \Sigma_k^{-1} \mu_k + \mu_k^T \Sigma_k^{-1} \mu_k  \Big] \Bigg\} 
    \\
    &= \mathbb{E} \Bigg\{\sum_i \mathbbm{1}(z_i = k) \Big[ (x_k - \mu_k)^T \Sigma_k^{-1}(x_k - \mu_k) + (x_i - \bar{x}_k)^T\Sigma_k^{-1}(x_i - \bar{x}_k)  \Big] \Bigg\}
    \\
    &= N_k \Big[(\bar{x}_k - m_k)^T v_k C_k^{-1} (\bar{x}_k - m_k) + \operatorname{tr}\big(\Sigma_k^{-1} \frac{\Sigma_k}{b_k}\big) + v_k \operatorname{tr}(S_k C_k^{-1}) \Big]
    \\ 
    &= N_k (\bar{x}_k - m_k)^T v_k C_k^{-1} (\bar{x}_k - m_k) + N_k d b_k^{-1} + N_k v_k \operatorname{tr}(S_k C_k^{-1}).
\end{align*}
Then, 
\begin{align*}
\mathbb{E}[\ln p(X|z, \mu, \Sigma)] &= \dfrac{1}{2} \sum_k N_k \Big\{-d \ln 2\pi - \mathbb{E} (\ln |\Sigma_k|)- d b_k^{-1} 
\\
& \hspace{24mm}- v_k (\bar{x}_k - m_k)^T C_k^{-1} (\bar{x}_k - m_k) - v_k Tr(S_k C_k^{-1}) \Big\}.
\end{align*}

Now, we consider $\mathbb{E}[\log p(z)]$
\begin{align*}
    \mathbb{E}[\log p(z)] & = \mathbb{E} \Big[ \sum_{i=1}^n \sum_{k=1}^K \mathbbm{1}\{z_i = k\} \ln \pi_k + \sum_{i=1}^n \sum_{k=1}^K \sum_{j \in N(i)} \sum_{h=1}^K \lambda^*_{kh} \mathbbm{1}\{z_i = k, z_j = h, z_i \neq z_j\} \Big]
    \\
    &= \sum_{i=1}^n \sum_{k=1}^K c_{ik} \ln \pi_k + \sum_{i=1}^n \sum_{k=1}^K \sum_{j \in N(i)} \sum_{h=1}^K \lambda^*_{kh} c_{ik} c_{i}.
\end{align*}

For $\mathbb{E}[\log p(\mu,\Sigma)]$, notice that the expectation calculation should be split into $2$ parts, one with existing clusters and another for the new cluster. We first consider the expectation of the log probability for existing clusters as follows 
\begin{align*}
    \mathbb{E}[\log p(\mu,\Sigma)]_{k \in [\tilde{K}]} &= \mathbb{E} \Bigg[ \dfrac{\tilde{\nu}^{\text{adj}}_k}{2} \ln |\tilde{\Psi}^{\text{adj}}_k| - \dfrac{\tilde{\nu}^{\text{adj}}_k + d + 1}{2} \ln |\Sigma_k| - \dfrac{1}{2} \operatorname{tr}(\tilde{\Psi}^{\text{adj}}_k \Sigma_k^{-1}) - \dfrac{d \tilde{\nu}^{\text{adj}}_k}{2} \ln 2 
    \\
    & \hspace{10mm} - \ln \Gamma \Big(\dfrac{\tilde{\nu}^{\text{adj}}_k}{2}\Big) + \dfrac{d}{2}\ln \Big(\frac{\tilde{\beta}_k}{2\pi}\Big) - \dfrac{1}{2} \ln|\Sigma_k| - \dfrac{\tilde{\beta}_k}{2} (\mu_k\!-\! \tilde{\eta}_k)^T \Sigma_k^{-1} (\mu_k\!-\! \tilde{\eta}_k)  \Bigg]
    \\
    & = \sum_{k=1}^{\tilde{K}} \Bigg[ - \dfrac{\tilde{\nu}^{\text{adj}}_k + d + 2}{2} \mathbb{E}(\ln |\Sigma_k|) - \dfrac{1}{2} v_k \operatorname{tr}(\Psi_k C_k^{-1}) + \dfrac{d}{2}\ln \Big(\frac{\tilde{\beta}_k}{2\pi}\Big) 
    \\
    & \hspace{15mm} - \dfrac{\tilde{\beta}_k}{2} (m_k - \tilde{\eta}_k)^T v_k C_k^{-1} (m_k - \tilde{\eta}_k)   
    \\
    &\hspace{15mm}- \dfrac{\tilde{\beta}_k}{2} \operatorname{tr}\Big(\Sigma_k^{-1} \operatorname{var}(\mu_k - \tilde{\eta}_k)(\mu_k - \tilde{\eta}_k)^T \Big) + \ln B(\tilde{\Psi}^{\text{adj}}_k, \tilde{\nu}^{\text{adj}}_k) \Bigg]
    \\ 
    & = \sum_{k=1}^{\tilde{K}} \Bigg[ - \dfrac{\tilde{\nu}^{\text{adj}}_k + d + 2}{2} \mathbb{E}(\ln |\Sigma_k|) - \dfrac{1}{2} v_k \operatorname{tr}(\tilde{\Psi}^{\text{adj}}_k C_k^{-1}) + \dfrac{d}{2}\ln \Big(\frac{\tilde{\beta}_k}{2\pi}\Big)
    \\
    & \hspace{15mm} - \dfrac{\tilde{\beta}_k}{2} (m_k - \tilde{\eta}_k)^T v_k C_k^{-1} (m_k - \tilde{\eta}_k) - \dfrac{d \tilde{\beta}_k}{2 b_k}  +  \ln B(\tilde{\Psi}^{\text{adj}}_k, \tilde{\nu}^{\text{adj}}_k) \Bigg],
\end{align*}

where $B(\cdot)$ is defined in (\ref{eq: function B}) and $\Gamma(\cdot)$ is the Gamma function.

With new clusters, the process for expectation computation remains unchanged except for the prior parameters, which are 
\begin{align*}
    \mathbb{E}[\log p(\mu,\Sigma)]_{k = \tilde{K} + 1,\ldots, K} & = \sum_{k=\tilde{K}+1}^{K}-\dfrac{\nu+ d + 2}{2} \mathbb{E}(\ln |\Sigma_k|) - \dfrac{1}{2} v_k \operatorname{tr}(\Psi C_k^{-1}) + \dfrac{d}{2}\ln \Big(\frac{\kappa}{2\pi}\Big) 
    \\
    & \hspace{8mm} - \dfrac{\kappa}{2} (m_k - \eta)^T v_k C_k^{-1} (m_k - \eta) - \dfrac{d \kappa}{2 b_k} +  \ln B(\Psi, \nu).
\end{align*}

Consider $\mathbb{E}[\log q(z)]$, we have
\[
\mathbb{E}[\log q(z)] = \mathbb{E}\Big(\sum_i \sum_k \mathbbm{1}\{z_i = k\}\ln c_{ik} \Big) = \sum_i \sum_k c_{ik} \ln c_{ik}.
\]

Finally, consider $\mathbb{E}[\log q(\mu,\Sigma)]$
\begin{align*}
    \mathbb{E}[\log q(\mu, \Sigma)] &= \mathbb{E} \Bigg[ \dfrac{\tilde{\nu}^{\text{adj}}_k}{2} \ln |\tilde{\Psi}_k| - \dfrac{\tilde{\nu}^{\text{adj}}_k + d + 2}{2} \ln |\Sigma_k| - \dfrac{1}{2} \operatorname{tr}(\tilde{\Psi}^{\text{adj}}_k \Sigma_k^{-1}) - \dfrac{\tilde{\nu}^{\text{adj}}_k d}{2} \ln 2 
    \\
    & \hspace{11mm} - \ln \Gamma \Big(\dfrac{\tilde{\nu}^{\text{adj}}_k}{2} \Big) + \dfrac{d}{2}\ln \Big(\frac{\tilde{\beta}_k }{2\pi} \Big) -\dfrac{\tilde{\beta}_k}{2} (\mu_k - \tilde{\eta}_k)^T \Sigma_k^{-1} (\mu_k - \tilde{\eta}_k)  \Bigg]
    \\
    &= \sum_{k=1}^K \Bigg\{\ln B(C_k, v_k) - \dfrac{\tilde{\nu}^{\text{adj}}_k + d + 1}{2} \mathbb{E}(\ln |\Sigma_k|) - \dfrac{v_kd}{2}
    \\
    &\hspace{11mm} - \dfrac{1}{2}\mathbb{E}(\ln |\Sigma_k|)  + \dfrac{d}{2}\ln \Big(\frac{b_k }{2\pi} \Big) - \dfrac{d}{2} \Bigg\}
    \\
    & = \sum_{k=1}^K \Bigg\{H\Big[q(\Sigma_k)\Big] - \dfrac{1}{2}\mathbb{E}(\ln |\Sigma_k|) + \dfrac{d}{2}\ln \Big(\frac{b_k}{2\pi} \Big) - \dfrac{d}{2} \Bigg\},
\end{align*}
where $H(\cdot)$ is defined in (\ref{eq: function H}).

\subsection{Algorithm Initialisation\label{sec:alg_init}}
We use the $\tilde{\eta}_k$ (estimated from the external data) to find suitable starting points $c_0$ for the CAVI algorithm. To this end, we calculate the distance from each pixel to each known cluster mean, i.e. 
\[
\tilde{d}_{ik} = \|x_i - \tilde{\eta}_k\|^2, \quad  \text{for} \quad 1 \leq k \leq \tilde{K}, 1\leq i\leq n,
\]
while for unknown clusters, we choose the distance as 
\[
\tilde{d}_{ik} = Q_{0.75}\big( \min_{k} \tilde{d}_{ik} \big)_{i=1}^{N}, \quad \text{for} \quad  \tilde{K} + 1\leq k \leq K,  1\leq i\leq n,
\]
where $Q_{0.75}$ denotes the third quartile. 

We introduce a scaling parameter $s^2$ and let
\[
d_{ik} = \dfrac{\exp(\tilde{d}_{ik} /s^2)}{\sum_h \exp(\tilde{d}_{ih}/s^2)}.
\]
Setting $s^2$ small leads to sharper cluster assignments, while large values result in more uncertain memberships. We add a noise term $\gamma \sim \mathcal{D}(\xi)$, where $\xi$ is determined by the user (we found $\xi=1/2$ to be a sensible choice) and introduce a weight parameter $w$ to control the amount of noise (for this parameter, we found $w=1/2$ to be a good choice in our experiments). We then let  
\[ 
\tilde{c}^{(0)}_{ik} = d_{ik} + w \cdot \gamma, 
\]
and normalise each row to yield the initialisation for the variational parameter $c$ as
\begin{equation} 
    c^{(0)}_{ik} = \dfrac{\tilde{c}^{(0)}_{ik}}{\sum_h \tilde{c}^{(0)}_{ih}} .
    \label{eq:c_0_initialisation}
\end{equation}

We initialise the other variational parameters based on values estimated from the external data as in (\ref{eq:external_data_posterior}), while for the unknown clusters, since we have no information about these clusters, we choose to initialise with non-informative priors, i.e.
\begin{align}\begin{split}
    m^{(0)}_k &= 
    \begin{cases}
        \tilde{\eta}_k \quad \text{if} \quad 1 \leq k \leq \tilde{K}
        \\
        \eta \quad \text{if} \quad \tilde{K} + 1 \leq k \leq K
    \end{cases}
    \\
    v^{(0)}_k &= 
    \begin{cases}
        \tilde{\nu}^{\text{adj}}_k \quad \text{if} \quad 1 \leq k \leq \tilde{K} 
        \\ 
        \nu \quad \text{if} \quad \tilde{K} + 1 \leq k \leq K
    \end{cases}
    \\
    C^{(0)}_k &= 
    \begin{cases}
        \tilde{\Psi}^{\text{adj}}_k \quad \text{if} \quad 1 \leq k \leq \tilde{K} 
        \\
        \Psi \quad \text{if} \quad \tilde{K} + 1 \leq k \leq K.
    \end{cases}
\end{split}\end{align}

\subsection{Algorithm for \texorpdfstring{$\zeta$}{zeta} selection} \label{sec:zeta_alg_appendix}

\begin{algorithm}[H]
\caption{Procedure to determine choice of $\zeta$\label{alg:zeta_alg}}
\begin{algorithmic}[1]
\setlength{\itemsep}{-0.5pt}
\State \textbf{Inputs:} 
\State \quad External data $\tilde{X}$, test data $X$, number of known clusters $\tilde{K}$.

\Statex
\State {\bf{Step 1}}
\State Run Algorithm~\ref{alg:potters}, with $\zeta_k = 0.01$ for all $k$.
\State Run Algorithm~\ref{alg:potters}, with $\zeta_k = 1$ for all $k$.
\State Run Algorithm~\ref{alg:potters}, with $\zeta_k = 100$ for all $k$.
\Statex
\State {\bf Step 2}
\State For each $k$, pick $\zeta_k \in \{0.01, 1, 100\}$ based on the results from Step~1. 
\State Run Algorithm~\ref{alg:potters} with the new choice for the trust parameters $\zeta_k$.
\Statex
\State {\bf Step 3}
\State Adjust $\zeta_k$ based on the results from Step~2 by diving or multiplying individual $\zeta_k$ by $10$ if neccessary.
\Statex
\State \textbf{Outputs:} The choice for $\zeta = (\zeta_1, \zeta_2, \cdots, \zeta_{\tilde{K}})$.
\end{algorithmic}
\end{algorithm}

\subsection{Hyperparameter Selection for the Simulation Study\label{sec:hyperparam_sim_data}}
We specify the hyperparameter choice in our simulation study here. For fixed hyperparameters, our choice is based on the knowledge about the data.
\begin{align*}
    \eta &= \dfrac{1}{N}\displaystyle \sum_{i=1}^N x_i
    \\
    \Psi &= \dfrac{1}{N-1}\displaystyle \sum_{i=1}^N \dfrac{(x_i - \eta)^2}{N-1}
    \\
    \nu &= D+2 
    \\
    \kappa &= 10^{-3}
    \\
    \zeta_k &= 10^2 \quad \text{for every} \quad k \in [K']
    \\
    \pi_k &= 
    \begin{cases}
        \dfrac{N_k}{2N} + \dfrac{1}{2K} \, & \text{for} \quad 1 \leq k \leq \tilde{K}  
        \\
        \\
        \dfrac{1}{2K} \, &\text{for} \quad \tilde{K} +1 \leq k \leq K.
    \end{cases}
\end{align*}
Given our strong confidence in the labelled data and prior knowledge that the target data contains additional clusters, we select a large $\zeta_k$ for every class, and set $\epsilon =0.5$ in the construction of $\pi$ to allow flexibility in identifying new clusters. The parameters $\tilde\eta$, $\tilde\kappa$, $\tilde\nu$, and $\tilde\Psi$ are then computed as presented in Equation~\eqref{eq:external_data_posterior};  the transformed parameters $\tilde\nu^{\text{adj}}$ and $\tilde\Psi'$ are then given by $\tilde{\nu}^{\text{adj}}_k = \dfrac{\zeta_k \tilde{\nu}^{\text{adj}}_k + d + 1}{\zeta_k + 1}$ and $\tilde{\Psi}'_k = \dfrac{\zeta_k}{\zeta_k + 1} \tilde{\Psi}^{\text{adj}}_k$. The $\widehat\Lambda^*$ are estimated from the external data as specified in Section~\ref{sec:hyperparameter_spec}.

\subsection{Data Preprocessing for the Case Study\label{sec:case_study_details}}
We describe the detailed steps used to preprocess the image data in the case study. For both the externally labelled dataset and the target data, we first define the geographic area of interest. For the external data, the area of interest is a region in the North East of England, specified by the bounding box 
$\text{[-2.46, 55.28, -2.32, 55.347]}$ with three images of this area in three seasons: Autumn (September~1,~2019-November~30,~2019); Spring (March~1,~2020 -May~31,~2020); Summer (June~1,~2020-August~31,~2020). We chose these time periods between 2019 and 2020 as the labels we are using were estimated from satellite imagery during this time. For the target area, we define the bounding box $\text{[-2.60, 56.85, -2.50, 56.95]}$, and collect three images corresponding to the same seasons as the external data but in different years: Autumn (September~1,~2023 -November~30,~2023); Spring (March~1,~2024-May~31,~2024); Summer (June~1,~2024-August~31,~2024). 

\begin{figure}
    \centering
    \includegraphics[width=\linewidth]{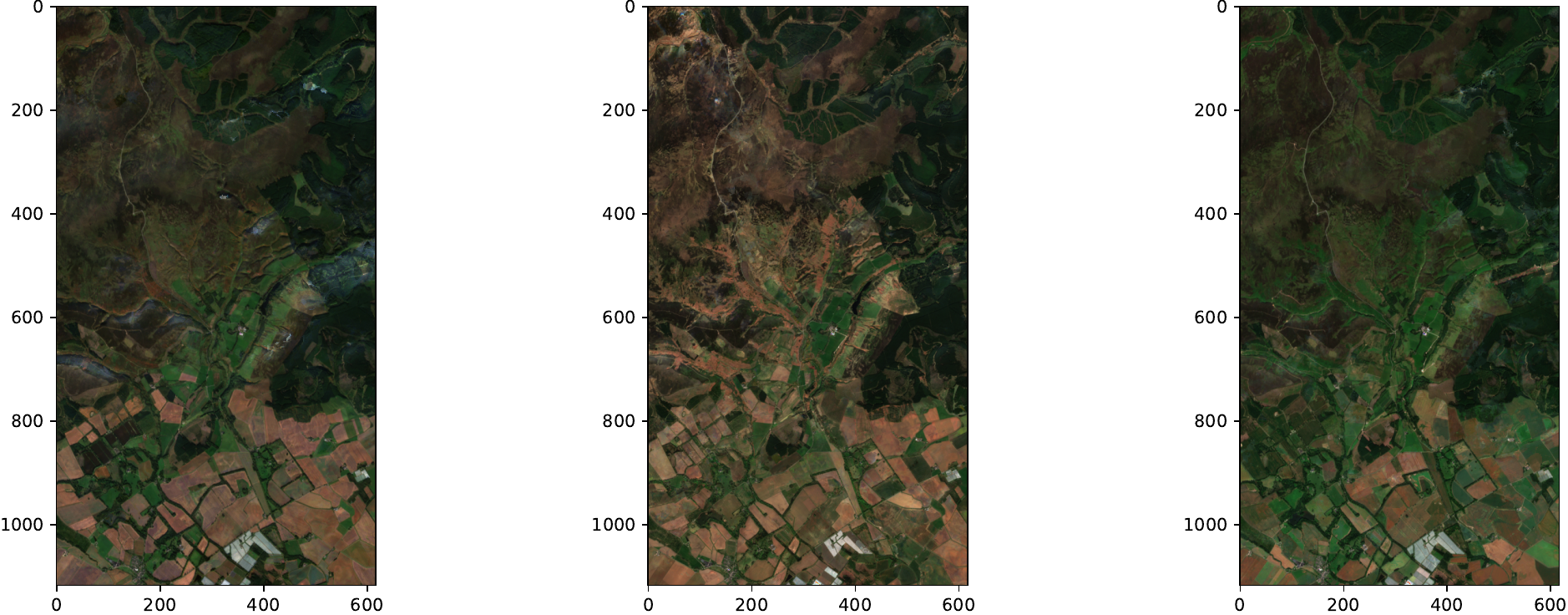}
    \caption{The satellite images of the target region near Glensaugh (Scotland) corresponding to three seasons: Autumn (September~1,~2023--November~30,~2023); Spring (March~1,~2024--May~31,~2024); Summer (June~1,~2024--August~31,~2024).}
    \label{fig:Glensaugh_orginal_data}
\end{figure}

\begin{figure}
    \centering
    \includegraphics[width=\linewidth]{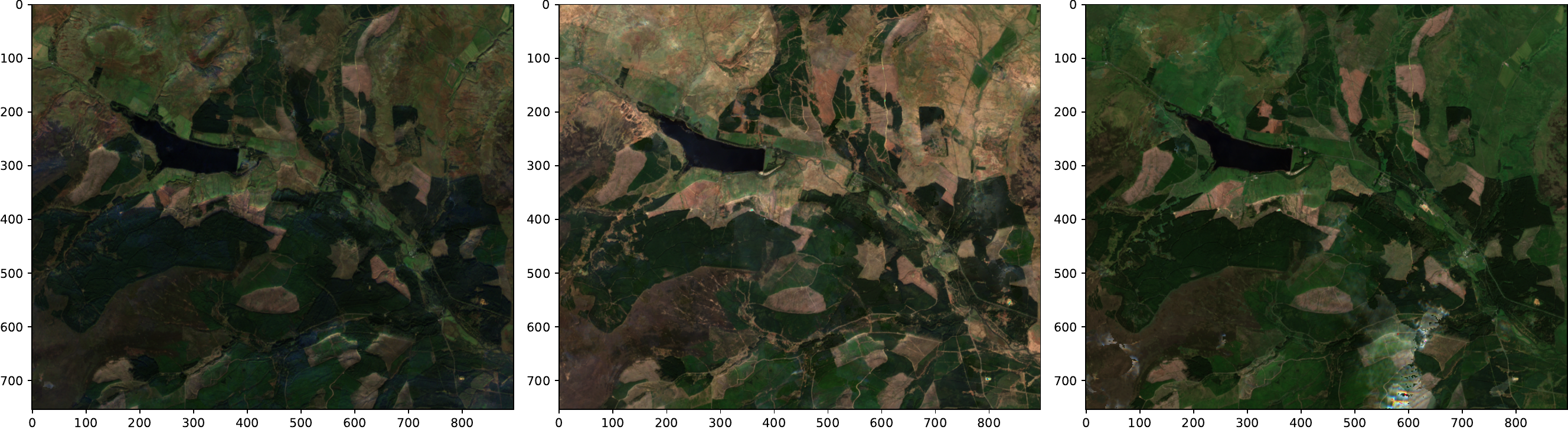}
    \caption{The RGB satellite image of the North-Eastern region in England as external data in three seasons: Autumn (September~1,~2019--November~30,~2019); Spring (March~1,~2020--May~31,~2020); Summer (June~1,~2020--August~31,~2020).}
    \label{fig:England_orginal_data}
\end{figure}

For each interested region and each time period, we access the Sentinel-2 satellite imagery database via the Earth Search STAC API (available at \url{lhttps://earth-search.aws.element84.com/v1ink}) using the \texttt{pystac-client} package in Python. We then selected the \texttt{sentinel-2-c1-l2a} collection from the available datasets. The colour channels used in the images include Red (R), Green (G), Blue (B), Near-Infrared (NIR), Shortwave Infrared 16 (SWIR16), and Shortwave Infrared 22 (SWIR22). Under some weather conditions, the `water' and `grass' areas are known to be hard to identify in the satellite images using these channels; we thus decided to include the Normalised Difference Water Index (NDWI) channel, which is calculated as the ratio of the difference and the sum of Green and NIR colour channels.  Due to the Scottish weather conditions, satellite images usually contain cloud cover, which reduces the efficiency of the algorithm. We only keep satellite images with a cloud cover percentage below 50\% and remove the Scene Classification Layer and missing pixels in the images. 

Since the Sentinel-2 satellite images are reflectance stored as integers with a scale given by a quantification value (usually $10,000$), we need to normalise the raw Sentinel-2 reflectance data using the appropriate scaling factor and offset to convert integer pixel values to physical reflectance values between 0 and 1. In our case, the data is normalised using a scaling factor of $0.0001$ and an offset of $0.1$, 
\[\text{normalised data} = \text{data} \times 0.0001 - 0.1.\] 
The details on how to choose different scaling parameters and calculate the normalised data in different cases can be found at \url{https://clearsky.vision/knowledge/sentinel2-scaling-harmonization}. 

From the collection of satellite images in each season, we generate a single image for each time period by computing the \emph{median} reflectance at each pixel across all available images, after having removed any pixels that are highly likely to contain clouds.  The external and target data are then created by stacking all channels for all seasons into a single dataframe, which is ready for further analysis.

For the external data, an additional step is required to process the label dataset. Each pixel in the satellite images is converted into a point geometry and stored in a GeoDataFrame, which is then spatially joined with the labelled polygons to assign class labels to corresponding pixels. The joined dataset is sorted by spatial coordinates, and unnecessary columns and pixels corresponding to missing labels are dropped. 

All analyses were performed on the resources provided by the Edinburgh Compute and Data Facility (ECDF) (\url{http://www.ecdf.ed.ac.uk/}), using Python 3.10. All code is publicly available at \url{https://github.com/bknguyen11/POTTERS/tree/main}.

\subsection{Additional Figures for the Case Study\label{sec:appendix_maps}}

\begin{figure}[H]
    \centering
    \includegraphics[width=\linewidth]{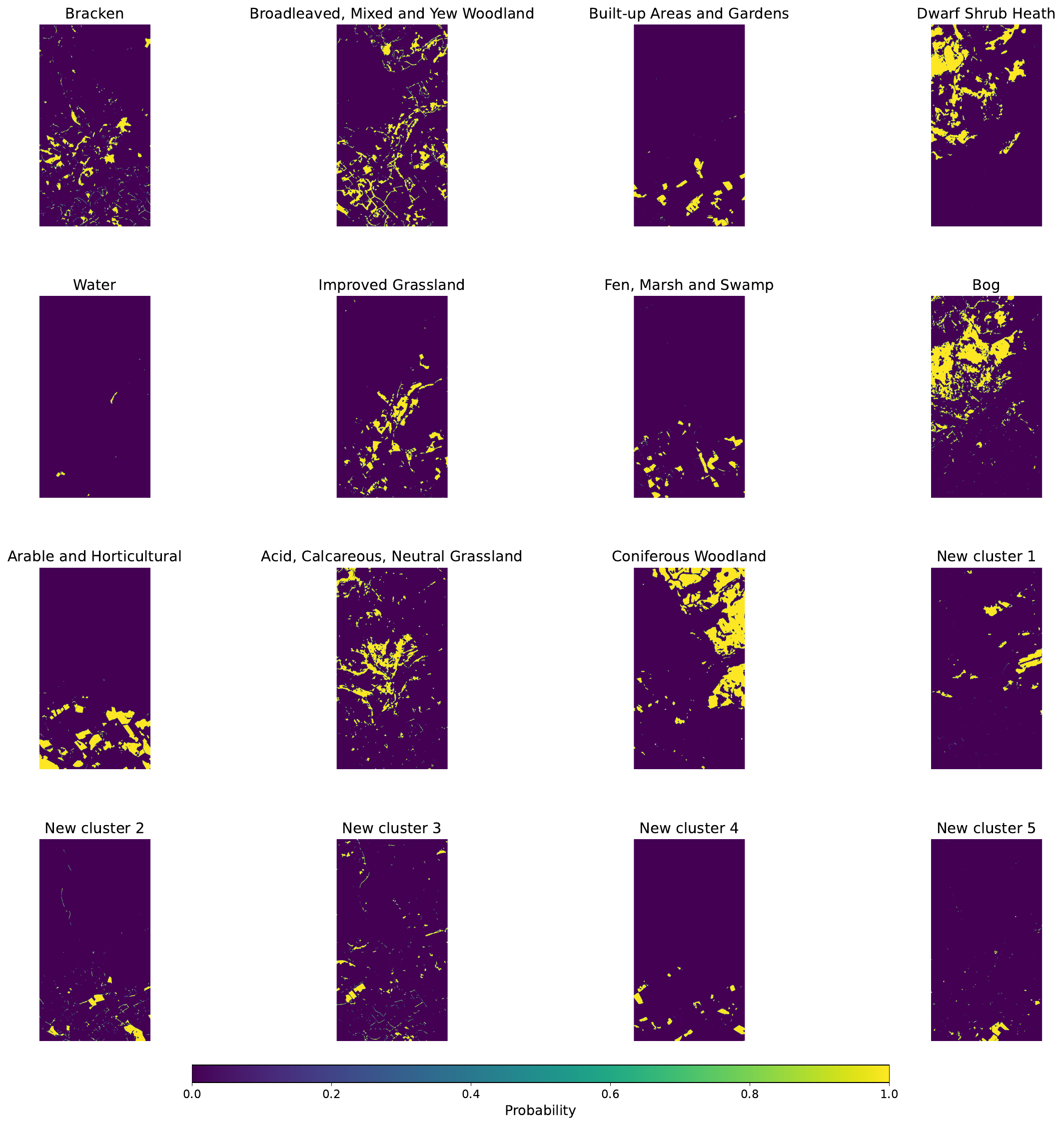}
    \caption{Probability maps for each cluster.}
    \label{fig:prob_maps}
\end{figure}

\begin{figure}
    \centering
    \includegraphics[width=1.00\linewidth]{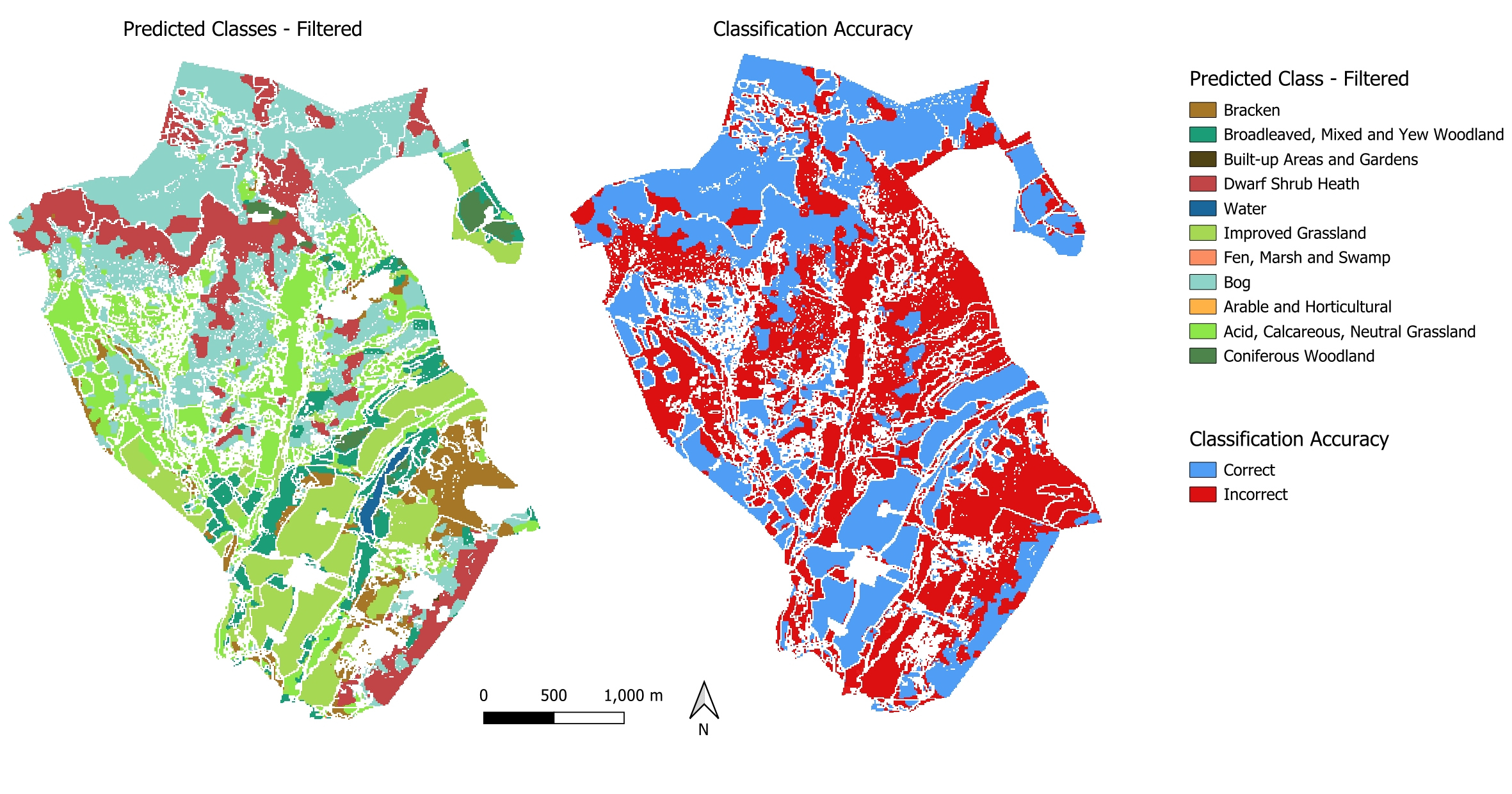}
    \caption{Accuracy map}
    \label{fig:validation_accuracy_map}
\end{figure}

\begin{figure}
    \centering
    \includegraphics[width=0.95\linewidth]{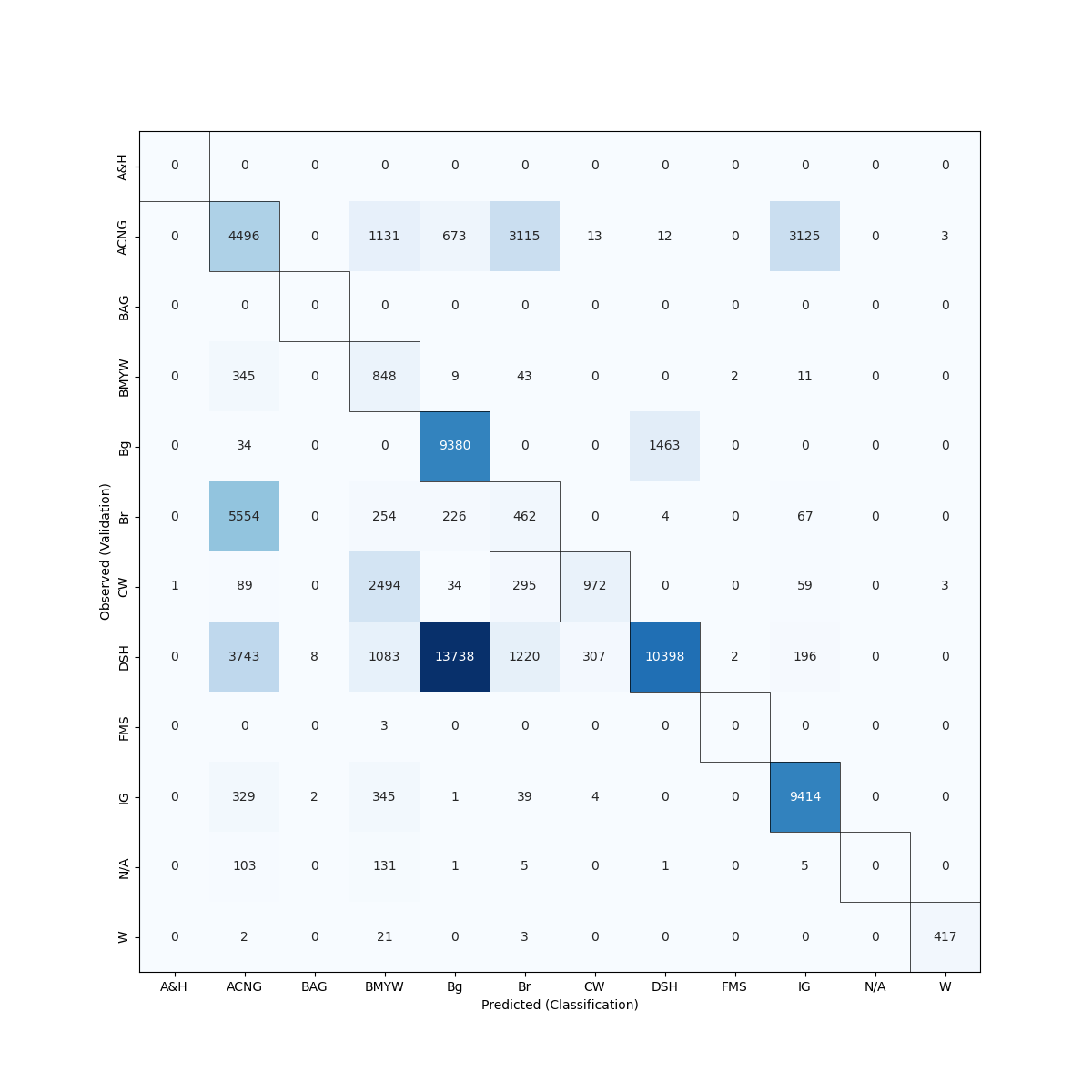}
    \caption{Confusion matrix}
    \label{fig:validation_confusion_matrix}
\end{figure}

\begin{figure}
    \centering
    \includegraphics[width=\linewidth]{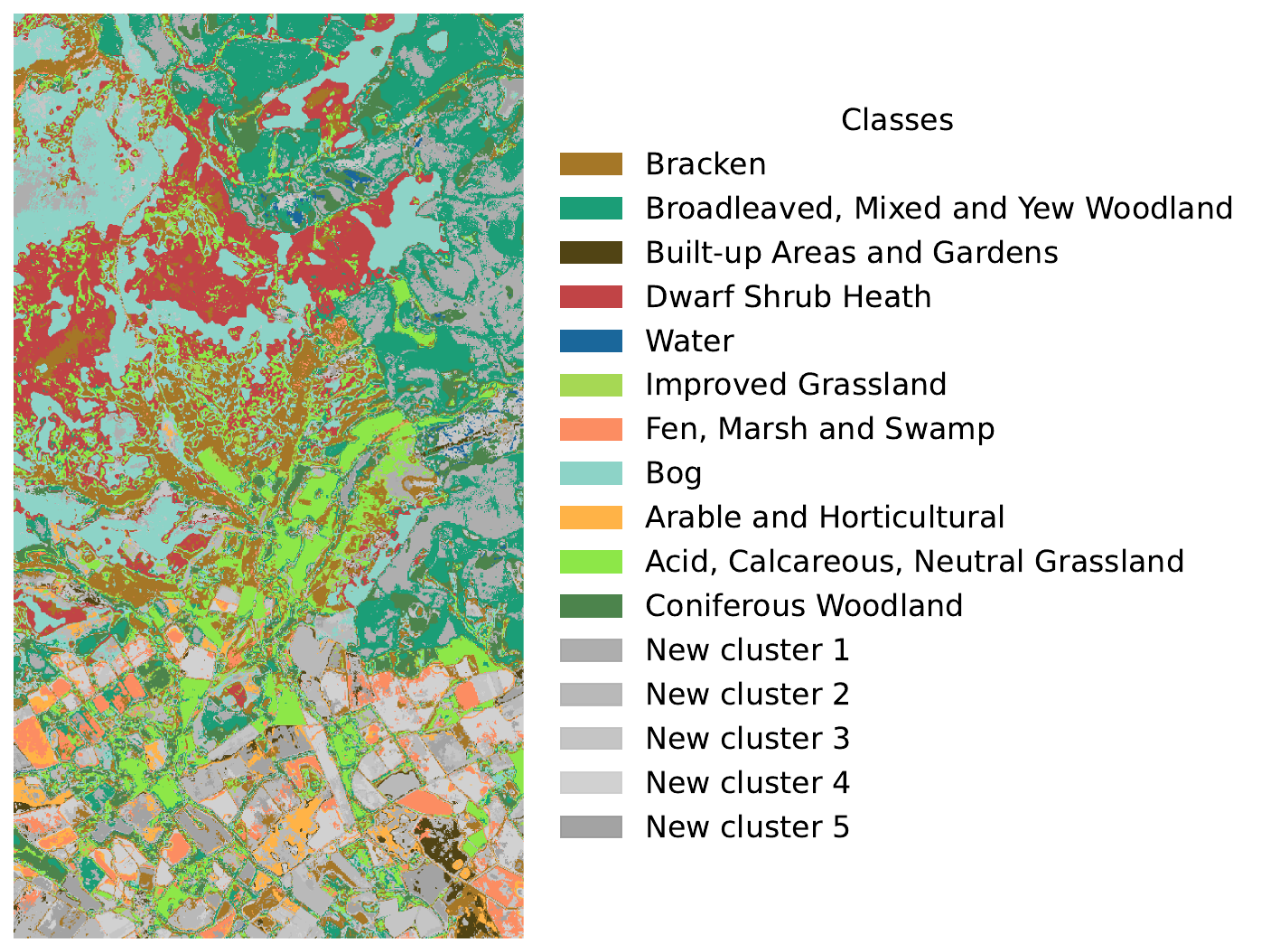}
    \caption{Maps for secondary-level classes\label{fig:Glensaugh_result_second_class}}
\end{figure}

\end{document}